\pdfoutput=1
\documentclass[12pt, authoryear]{elsarticle}
\usepackage{geometry} 
\usepackage{graphics}
\usepackage{amsmath} 
\usepackage{amssymb}
\usepackage{rotating}
\usepackage{subfigure}
\usepackage{hyperref}
\usepackage{longtable}
\usepackage{lscape}
\usepackage{setspace}

\title{Asteroid taxonomic signatures from photometric phase curves}
\author{
D.~A.~Oszkiewicz$^{1,2,3}$, E.~Bowell$^2$, L.~H.~Wasserman$^2$, \\
 K.~Muinonen$^{1,4}$, A.~Penttil\"{a}$^1$, T.~Pieniluoma$^1$, \\
  D.~E.~Trilling$^3$, C.~A.~Thomas$^3$ \\
\small{ $^1$ Department of Physics, P.O. Box 64, FI-00014 University of Helsinki, Finland.} \\
\small{ $^2$ Lowell Observatory, 1400 West Mars Hill Road, Flagstaff, AZ 86001, U.S.A.} \\
\small{ $^3$ Department of Physics and Astronomy, Northern Arizona University,} \\ 
\small{ P.O. Box 6010, Flagstaff, AZ 86011, U.S.A.} \\
\small{ $^4$ Finnish Geodetic Institute, P.O. Box 15, FI-02431 Masala, Finland.}
}
\date{\today} 

\begin{document}

\doublespacing

\begin{abstract}

  We explore the correlation between an asteroid's taxonomy and
  photometric phase curve using the $H,\!G_{12}$ photometric phase
  function, with the shape of the phase function described by the
  single parameter $G_{12}$.  We explore the usability of $G_{12}$ in
  taxonomic classification for individual objects, asteroid families,
  and dynamical groups. We conclude that the mean values of $G_{12}$
  for the considered taxonomic complexes are statistically different,
  and also discuss the overall shape of the $G_{12}$ distribution for
  each taxonomic complex. Based on the values of $G_{12}$ for about
  half a million asteroids, we compute the probabilities of C, S, and
  X complex membership for each asteroid. For an individual asteroid,
  these probabilities are rather evenly distributed over all of the
  complexes, thus preventing meaningful classification.  We then
  present and discuss the $G_{12}$ distributions for asteroid
  families, and predict the taxonomic complex preponderance for
  asteroid families given the distribution of $G_{12}$ in each family.
  For certain asteroid families, the probabilistic prediction of
  taxonomic complex preponderance can clearly be made.  In particular,
  the C complex preponderant families are the easiest to detect, the
  Dora and Themis families being prime examples of such families.  We
  continue by presenting the $G_{12}$-based distribution of taxonomic
  complexes throughout the main asteroid belt in the proper element
  phase space. The Nysa-Polana family shows two distinct regions in
  the proper element space with different $G_{12}$ values dominating
  in each region. We conclude that the $G_{12}$-based probabilistic
  distribution of taxonomic complexes through the main belt agrees
  with the general view of C complex asteroid proportion increasing
  towards the outer belt. We conclude that the $G_{12}$ photometric
  parameter cannot be used in determining taxonomic complex for
  individual asteroids, but it can be utilized in the statistical
  treatment of asteroid families and different regions of the main
  asteroid belt.

\end{abstract}

\maketitle

\section{Introduction}

The photometric phase function describes the relationship between the
reduced magnitude (apparent magnitude at $1$~AU distance) and the
solar phase angle (Sun-asteroid-observer angle). Previously in
\cite{DO}, we have fitted $H,\!G_1,\!G_2$ and $H,\!G_{12}$ phase
functions presented in \cite{KM2010} for about half a million
asteroids contained in the Lowell Observatory database and obtained
absolute magnitudes and photometric parameter(s) for each
asteroid. The absolute magnitude $H$ for an asteroid is defined as the
apparent $V$ band magnitude that the object would have if it were
$1$~AU from both the Sun and the observer and at zero phase angle. The
absolute magnitude relates directly to asteroid size and geometric
albedo. The geometric albedo of an object is the ratio of its actual
brightness at zero phase angle to that of an idealized Lambertian disk
having the same cross-section. 

The shape of the phase curve described by the $G_1,\!G_2$ and $G_{12}$
parameters relates to the physical properties of an asteroid's
surface, such as geometric albedo, composition, porosity, roughness,
and grain size distribution. For phase angles larger than 10$^\circ$,
steep phase curves are characteristic of low-albedo objects with an
exposed regolith, whereas flat phase curves can indicate, for example,
a high-albedo object with a substantial amount of multiple scattering
in its regolith.

At small phase angles, atmosphereless bodies (such as asteroids)
exhibit a pronounced nonlinear surge in apparent brightness known as
the opposition effect \citep{KM2010a}. The opposition effect was first
recognized for asteroid (20) Massalia \citep{Gehrels}.  The
explanation of the opposition effect is two-fold: (1) self-shadowing
arising in a rough and porous regolith, and (2) coherent
backscattering; that is, constructive interference between two
electromagnetic wave components propagating in opposite directions in
the random medium \citep{KM2010a}. The width and height of the
opposition surge can suggest, for example, the compaction state of the
regolith and the distribution of particle sizes.

\cite{BelskayaShevchenko} have analyzed the opposition behavior of 33
asteroids having well-measured photometric phase curves and concluded
that the surface albedo is the main factor influencing the amplitude
and width of the opposition effect. Phase curves of high-albedo
asteroids have been also described by \cite{Harrisalbedo} and
\cite{ScaltritiZappala}. \cite{Harrisalbedo} have concluded that the
opposition spikes of (44)~Nysa and (64)~Angelina can be explained as
an ordinary property of moderate-to-high albedo atmosphereless
surfaces. \cite{KaasalainenS} have presented a method for interpreting
asteroid phase curves, based on empirical modeling and laboratory
measurements, and emphasized that more effort could be put into
laboratory studies to find a stronger connection between phase curves
and surface characteristics. Laboratory measurements of meteorite
phase curves have been performed, for example, by \cite{Capaccioni}
and measurements of regolith samples by \cite{KaasalainenSanna}.

The relationship between the phase-curve shape and taxonomy has also
been explored. \cite{LagerkvistMagnusson} have computed absolute
magnitudes and parameters for 69 asteroids using the $H,\!G$ magnitude
system and computed mean values of the $G$ parameter for taxonomic
classes S, M, and C. They have emphasized that the $G$ parameter varies
with taxonomic class. \cite{GoidetDevel} have considered phase curves
of about 35 individual asteroids and analogies between phase curves of
asteroids belonging to different taxonomic classes. \cite{HarrisTaxa}
have examined the mean values of slope parameters for different
taxonomic classes.

In our previous study \citep{DO}, we have fitted phase curves of about
half a million asteroids using recalibrated data from the Minor Planet
Center\footnote{IAU Minor Planet Center, see
  http://minorplanetcenter.net/iau/mpc.html}.  We have found a
relationship between the family-derived photometric parameters $G_1$
and $G_2$, and the median family albedo. We have showed that, in
general, asteroids in families tend to have similar photometric
parameters, which could in turn mean similar surface properties. We
have also noticed a correlation between the photometric parameters and
the Sloan Digital Sky Survey color indices (SDSS).  The SDSS color
indices correlate with the taxonomy, as do the photometric parameters.

In the present article, we explore the correlation of the photometric
parameter $G_{12}$ with different taxonomic complexes. For about half
a million individual asteroids, we compute the probabilities of C, S,
and X complex membership given the distributions of their $G_{12}$
values.  Based on the $G_{12}$ distributions for members of asteroid
families, we investigate taxonomic preponderance in asteroid
families. In Sec.~\ref{methods}, we describe methods to compute
$G_{12}$ for individual asteroids and the probability for an asteroid
to belong to a taxonomic complex given $G_{12}$, and methods to
determine the taxonomic preponderance in asteroid families. In
Sec.~\ref{res}, we describe our results and discuss the usability of
$G_{12}$ in taxonomic classification.  In Sec.~\ref{concl}, we present
our conclusions.

\section{Taxonomy from photometric phase curves}
\label{methods}

\subsection{Fitting phase curves}

In the previous study \citep{DO}, we made use of three photometric
phase functions: the $H,\!G$; the $H,\!G_1,\!G_2$; and the
$H,\!G_{12}$ phase functions.  The $H,\!G$ phase function \citep{EB}
was adopted by the International Astronomical Union in 1985. It is
based on trigonometric functions and fits the vast majority of the
asteroid phase curves in a satisfactory way. However, it fails to
describe, for example, the opposition brightening for E class
asteroids and the linear magnitude-phase relationship for F class
asteroids. The $H,\!G_{1},\!G_{2}$ and the $H,\!G_{12}$ phase
functions \citep{KM2010} are based on cubic splines and accurately fit
phase curves of all asteroids. The $H,\!G_{1},\!G_{2}$ phase function
is designed to fit asteroid phase curves containing large numbers of
accurate observations, whereas the $H,\!G_{12}$ phase function is
applicable to asteroids that have sparse or low-accuracy photometric
data. Therefore, the $H,\!G_{12}$ phase function is best suited to our
data \citep{DO}. The present study is mostly based on results obtained
from the $H,\!G_{12}$ phase function. Both $H,\!G_1,\!G_2$ and
$H,\!G_{12}$ phase functions are briefly described below.

\subsubsection{$H,\!G_1,\!G_2$ phase function}

In the $H,\!G_1,\!G_2$ phase function \citep{KM2010}, the reduced
magnitudes $V(\alpha)$ can be obtained from
\begin{equation}\begin{split}
    10^{-0.4 V(\alpha)} & =  a_1 \Phi_1(\alpha) + a_2 \Phi_2(\alpha) + a_3 \Phi_3(\alpha) \\
    & = 10^{-0.4H}\left[G_1 \Phi_1(\alpha) + G_2 \Phi_2(\alpha) +
      (1-G_1-G_2) \Phi_3(\alpha)\right],
\end{split}\end{equation}
where $\alpha$ is the phase angle and $V(\alpha)$ is the reduced
magnitude. The coefficients $a_1$, $a_2$, $a_3$ are estimated from the
observations using the linear least-squares method.  The basis
functions $\Phi_1(\alpha)$, $\Phi_2(\alpha)$, and $\Phi_3(\alpha)$ are
given in terms of cubic splines. The absolute magnitude $H$ and the
photometric parameters $G_1$ and $G_2$ can then be obtained from the
$a_1$, $a_2$, $a_3$ coefficients.

\subsubsection{$H,\!G_{12}$ phase function}

In the $H,\!G_{12}$ phase function \citep{KM2010}, the parameters
$G_1$ and $G_2$ of the three-parameter phase function are replaced by
a single parameter $G_{12}$ analogous to the parameter $G$ in the
$H,\!G$ magnitude system (although there is no exact
correspondence). The reduced flux densities can be obtained from
\begin{equation}
	10^{-0.4V(\alpha)} = L_0\left[G_1 \Phi_1(\alpha) + G_2\Phi_2(\alpha)+ 
	(1-G_1-G_2)\Phi_3(\alpha) \right],
\end{equation}
where
\begin{equation}\begin{split}
G_1 & = \begin{cases}
 0.7527 G_{12} + 0.06164,  & \mathrm{if} \ G_{12} < 0.2 \\
 0.9529 G_{12} + 0.02162, & \mathrm{otherwise}
\end{cases} \\
G_2 & = \begin{cases}
 -0.9612 G_{12} + 0.6270, & \mathrm{if} \  G_{12} < 0.2 \\
 -0.6125 G_{12} + 0.5572, & \mathrm{otherwise}
\end{cases}\end{split}
\label{G1G2}
\end{equation}
and $L_0$ is the disk-integrated brightness at zero phase angle. The
basis functions are as in the $H,\!G_1,\!G_2$ phase function. The
parameters $L_0$ and $G_{12}$ are estimated from the observations
using the nonlinear least-squares method.

The $H,\!G_1,\!G_2$ and the $H,\!G$ phase functions are fitted to the
observations in the flux-density domain using the linear least-squares
method. In order to fit the $H,\!G_{12}$ phase function, downhill
simplex non-linear regression \citep{NM} is utilized. in order to
compute uncertainties in the photometric parameters, we use Monte
Carlo and Markov-chain Monte Carlo methods. A detailed description of
these procedures can be found in \cite{DO}. In the current study, we
used the Asteroid Phase Curve Analyzer\footnote{Asteroid Phase Curve
  Analyzer --- an online java applet, available at
  \url{http://asteroid.astro.helsinki.fi/astphase/}}.

\subsection{Taxonomic preponderance for asteroid families}
\label{method}

\begin{figure}[ht]
\centering
\subfigure[C complex]{
  \includegraphics[width=0.45 \textwidth]{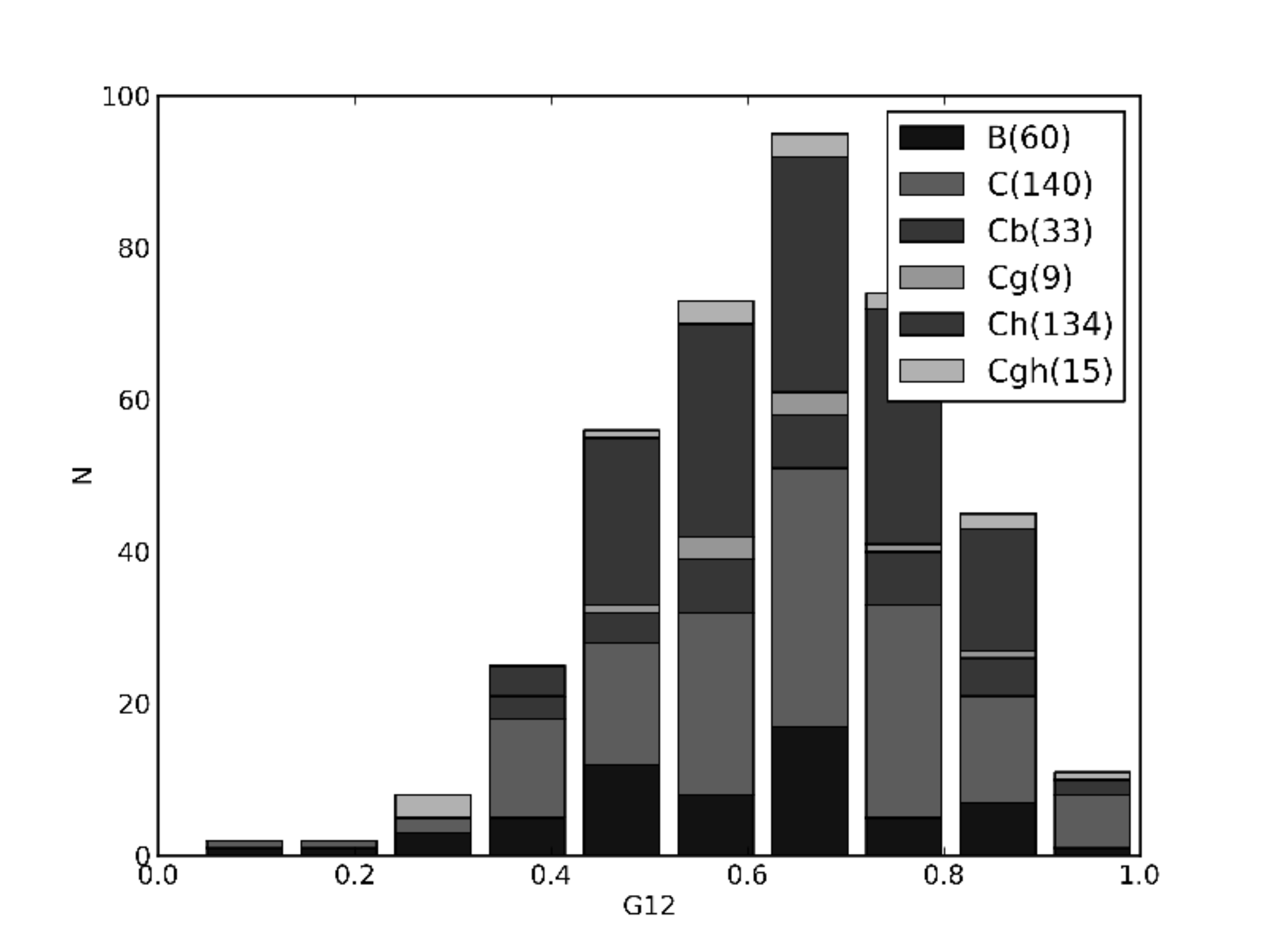} 
 }
 \subfigure[S complex]{
   \includegraphics[width=0.45 \textwidth]{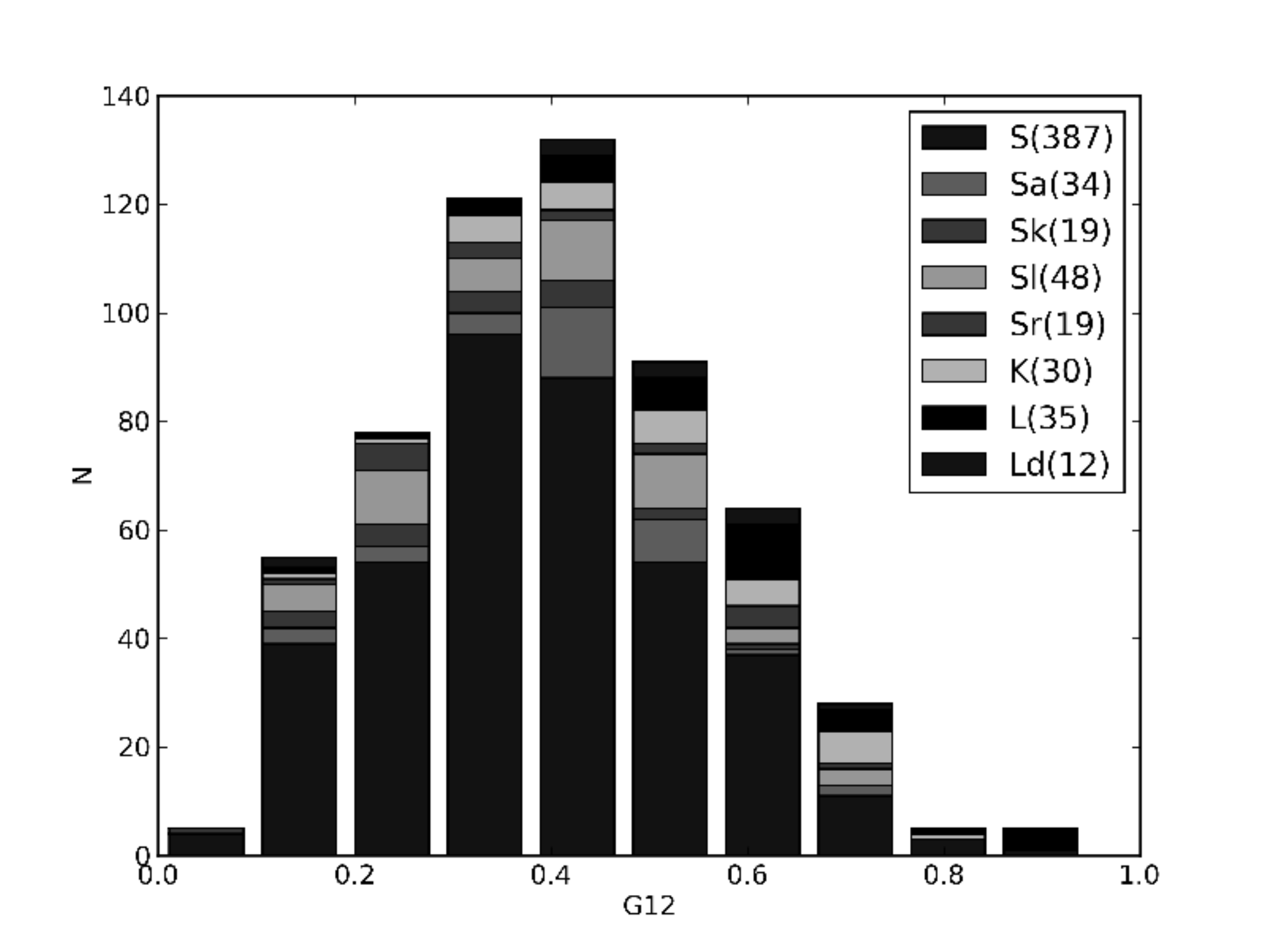}
 }
 \subfigure[X complex]{
   \includegraphics[width=0.45 \textwidth]{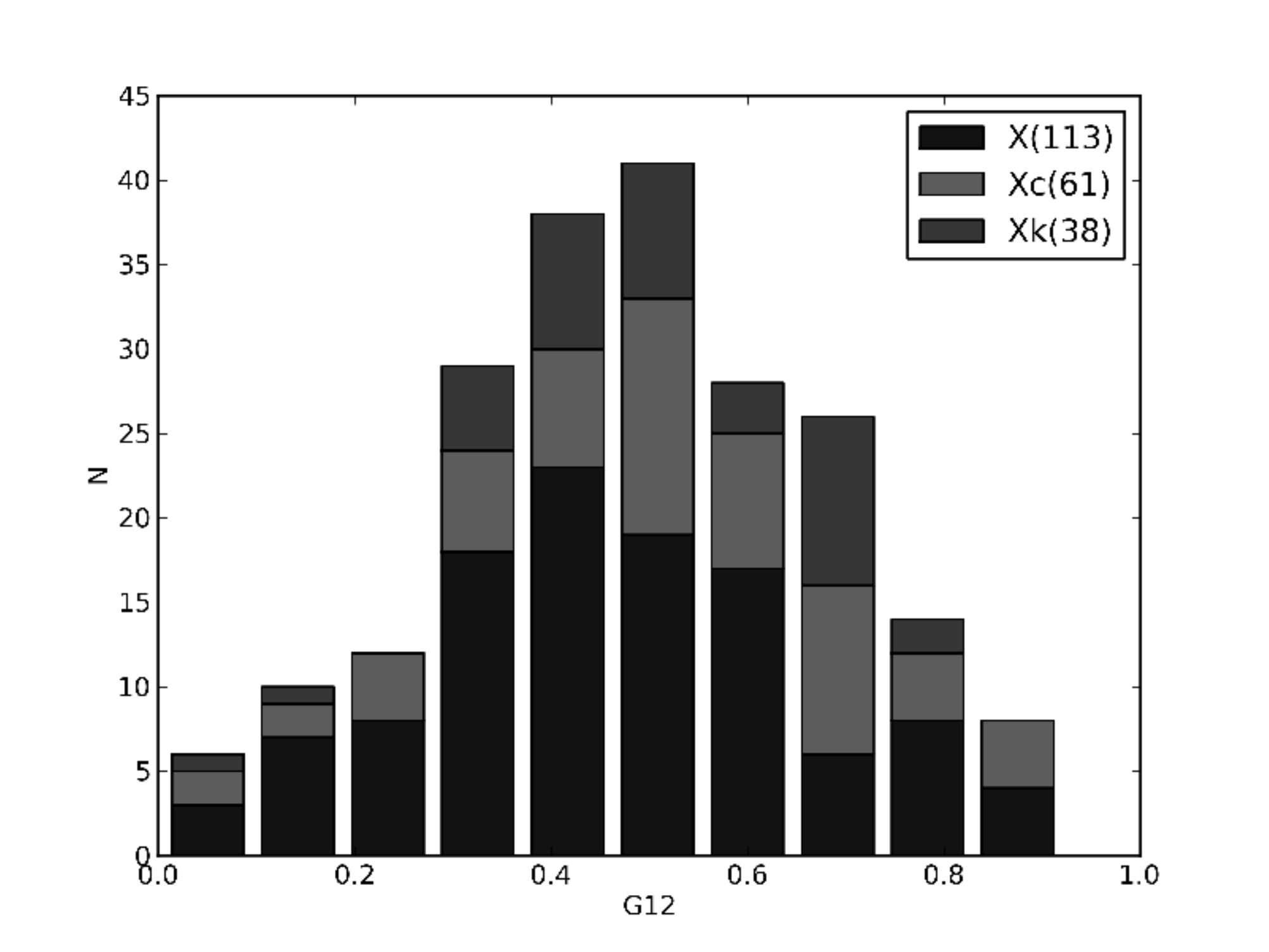}
 }
 \subfigure[Other complexes]{
  \includegraphics[width=0.45 \textwidth]{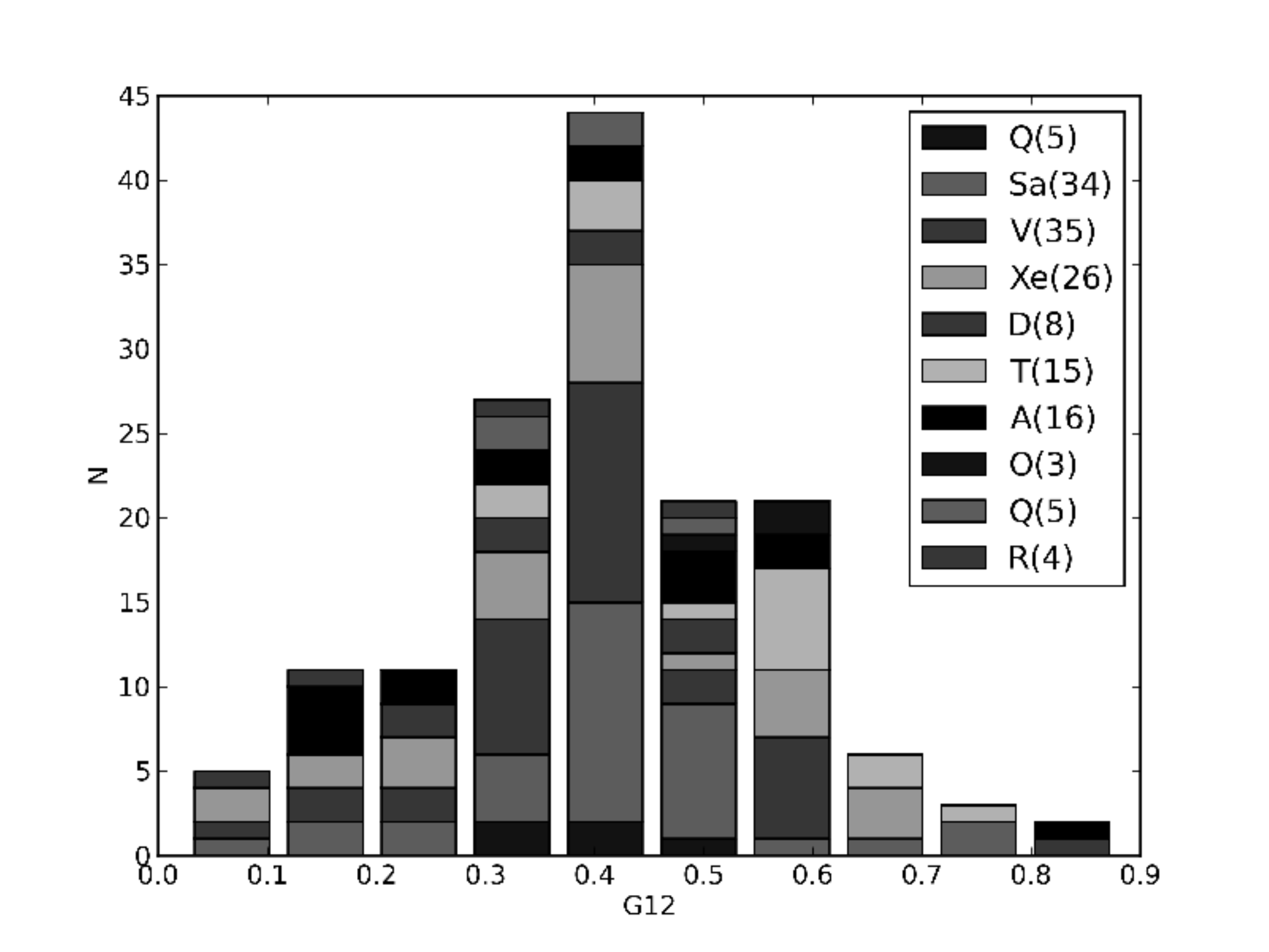}
 }
 \caption{Distributions of $G_{12}$ in asteroid taxonomic
   complexes. The different taxonomic classes are stacked from bottom
   to top in the histograms in the order listed in the legend.}
\label{taxaHisto}
\end{figure}

As we discuss further in Sec.~\ref{res}, we find correlation between
$G_{12}$ and taxonomy. In Fig.~\ref{taxaHisto}, we show $G_{12}$
histograms for different taxonomic complexes. Only the main taxonomic
complexes---that is, the C [containing classes B, C, Cb, Cg, Ch, Cgh],
S [S, Sa, Sk, Sl, Sr, K, L, Ld], and X [X, Xc, Xk] complexes---have
large enough sample size for statistical treatment.  The small number
of objects belonging to A, D [D, T], E [E, Xe], O, Q [Q, Sq], R, and V
complexes prevent further conclusions using $G_{12}$ statistics in
those groups. We approximate the $G_{12}$ distributions for taxonomic
complexes by a Gaussian distribution, and the means and standard
deviations of those distributions are listed in Table~\ref{G12taxa}.
The $G_{12}$ distributions for C and S complexes are smoother than the
one for the X complex. The X complex comprises three different albedo
groups, namely E, M, and P class objects. Those cannot be separated
within the X complex only based on spectra, and additional albedo
information is usually required. The X complex degeneracy was
discussed for example by \cite{CT2011}. The unusual shape of the X
complex can be related to the different albedo groups as $G_{12}$
correlates well with albedo \citep{KM2010, DO}. Unfortunately, due to
the small number of E, M, and P class objects in our sample, we
cannot determine how useful $G_{12}$ could be in breaking the X
complex into E, M, P class groups.

\begin{table}[htb]
\centering
\caption{Means and standard deviations of $G_{12}$ for asteroid taxonomic complexes.}
\label{G12taxa}
\vspace{1ex}
\begin{tabular}{llll} 
Complex & Nr of objects & mean & std \\ \hline
A & 16  &  0.39 & 0.19 \\
{\bf C} & {\bf 391}  &  {\bf 0.64} & {\bf 0.16} \\
D & 23  &  0.47 & 0.14 \\
E & 26  &  0.39 & 0.16 \\
O & 3  &  0.57 & 0.05 \\
Q & 72  &  0.41 & 0.14 \\
R & 4  &  0.24 & 0.18 \\
{\bf S} & {\bf 584}  &  {\bf 0.41} & {\bf 0.16} \\
V & 35  &  0.41 & 0.14 \\
{\bf X} & {\bf 212}  &  {\bf 0.48} & {\bf 0.19} \\  
\hline
\end{tabular}
\end{table} 

Based on the approximated $G_{12}$ distributions for the different
taxonomic complexes (Table \ref{G12taxa}), we can compute the
probability for an asteroid to belong to a given taxonomic complex as
the a posteriori probability using Bayes's rule.  For example, the
probability for an asteroid to belong to the C complex can be computed
using
\begin{equation}
p_C(x \in C \mid G_{12}) = A \,  pr(x \in C) \, p(G_{12} \mid x \in C),
\label{prob}
\end{equation}
where $A$ is a normalization constant, $p_C(x \in C \mid G_{12})$ is
the a posteriori probability for an asteroid $x$ to belong to the C
complex, given a particular $G_{12}$ value; $pr(x \in C)$ is the a
priori probability for an asteroid $x$ to belong to the C complex; and
$p(G_{12} \mid x \in C)$ is the probability for an asteroid $x$ to
have a specific $G_{12}$ value, given that it belongs to the C
complex. 

As estimates for the probabilities $p(G_{12} \mid x \in C)$, we adopt
the Gaussian approximations for the empirical $G_{12}$ distributions
of different taxonomic complexes. We make use of three different a
priori distributions: (1) a uniform a priori distribution; (2) an a
priori distribution based on the frequency of C, S, and X complex
objects among asteroids with taxonomy defined in the Planetary Data
System database \citep[PDS, see][]{pds}; (3) an a priori distribution
based on the frequencies of C, S, and, X complexes in different parts
of the main asteroid belt (inner, mid, and outer main belt) from the
PDS database.  Testing the results obtained with different a priori
distributions is important for insuring that the results are driven by
data and not by the a priori distribution.

The probabilities computed based on the different a priori
distributions should agree for different a priori assumptions if the
photometric parameter brings substantial information overriding the
information contained in the different a priori distributions.  By
using choice (1), we assume no previous knowledge of asteroid
taxonomy. By using choice (2), we assume that the a priori probability
for an asteroid $x$ to belong to a specific complex is equal to the
frequency of occurrence of asteroids of that complex in the sample of
known asteroid taxonomies in the PDS database. This means that, for
the C, S, and X complexes, we use the a priori probabilities equal to
$0.33$, $0.49$, and $0.18$, respectively.  To derive the a priori
distribution for choice (3), we first divide the main asteroid belt
into three regions: the inner (region I), mid (region II), and outer
main belt (region III). The boundaries between the regions are based
on the most prominent Kirkwood gaps. Region I lies between the 4:1
resonance ($2.06$~AU) and 3:1 resonance ($2.5$~AU). Region II
continues from the end of region I out to the 5:2 resonance
($2.82$~AU). Region III extends from the outer edge of region II to
the 2:1 resonance ($3.28$ AU). The frequencies derived for those
regions are as follows: for the C complex, 0.19 (I), 0.38 (II), 0.45
(III); for the S complex, 0.70 (I), 0.42 (II), 0.33 (III); and, for
the X complex, 0.11 (I), 0.20 (II), 0.22 (III). The regional
frequencies are also computed based on the data available in the PDS
database. In general, a better choice of the a priori distributions
would be based on debiased ratios of taxonomic complexes, but those
are not available. In general, a single asteroid can have non-zero
probability for belonging to two or more complexes.

The probability for an asteroid family being dominated, for example,
by the C complex can be computed as
\begin{equation}
P_{C} = \frac{\sum_{i=1}^{N_{mem}} p^{(i)}_C}{N_C+ N_S + N_X}, 
\end{equation}
where $N_{C}$, $N_{S}$, and $N_{X}$ are the numbers of asteroids
classified as belonging to the C, S, and X complexes, $N_{mem}$ is the
number of members in a family, and $p^{(i)}_C$ is the probability of
member $i$ belonging to the C complex. The probabilities for an
asteroid family being dominated by other complexes can be computed in
a similar fashion. In practice, $P_{C}$ represents the probability
that a random asteroid from a given family would be of C complex.

\subsection{Validation}

In order to validate the method described in Sec.~\ref{method}, we
have checked the number of correct taxonomic complex classifications
of asteroids with known taxa via so-called N-folded tests. First, we
derived the frequencies of different taxonomic complexes, skipping 50
random asteroids in each complex which we later use for testing. The
general frequencies for the C, S, and X complexes were $0.33$, $0.52$,
and $0.15$. In the inner, mid, and outer main belt, the numbers are,
respectively, as follows: 0.20, 0.72, and 0.09; 0.37, 0.44, and 0.18;
0.46, 0.37 and 0.17. Those frequencies are then used as priors in
Eq.~\ref{prob}.

The success ratio is measured as
\begin{equation}
R_s = \frac{N_{corr}}{N_{total}},
\end{equation}
where $N_{corr}$ is the number of correct identifications among
$N_{total}$ asteroids.

Using the uniform a priori distribution (1) results in a $64\%$
overall success ratio ($80\%$ for the C complex, $60\%$ for the S
complex, and $52\%$ for the X complex). Using the overall frequencies
(2) results in a $63\%$ overall success ratio ($98\%$ for the C
complex, $68\%$ for the S complex and $22\%$ for the X complex). The
last choice (3) leads to a $63 \%$ overall success ratio ($96\%$ for
the C complex, $72\%$ for the S complex and $22\%$ for the X
complex). We compared these success ratios with those arising from
random guessing. We conclude that there is general improvement in
success ratios for all the taxonomic complexes.

\section{Results and discussion}
\label{res}

%
%
%

\begin{figure}[htb]
   \centering
   \includegraphics[width=\textwidth]{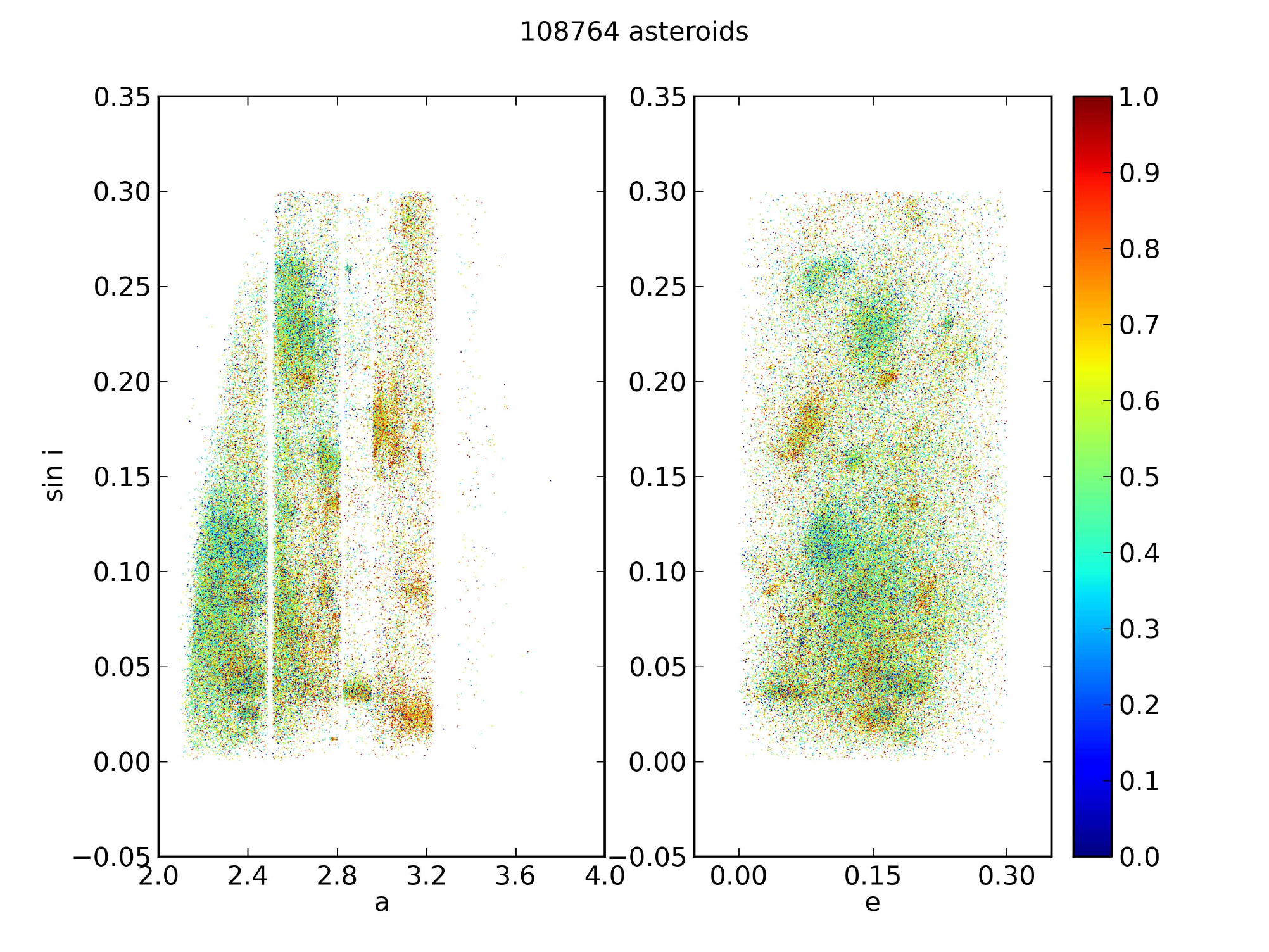} 
      \caption{Distribution of asteroid proper elements, color-coded
        according to the $G_{12}$ value.}
   \label{proper}
\end{figure}

We explore the correlation of the photometric parameter $G_{12}$ with
the taxonomic classification based on about half a million asteroid
phase curves in the Lowell Observatory database \citep{DO}.  In
Fig.~\ref{proper}, we present the distribution of the orbital proper
elements color-coded with the $G_{12}$ values, with with a larger
number of asteroids included as compared to the results in
\cite{DO}. The updated figure strengthens our previous findings of
$G_{12}$ homogeneity within asteroid families. Even though the
distributions of the $G_{12}$ values in asteroid families can be
broad, asteroids in families stand out and tend to have similar values
of $G_{12}$ \citep{DO}. Asteroids having disparate $G_{12}$ values but
still identified as family members can be so-called interlopers,
asteroids originating from a differentiated parent body, and asteroids
with differently evolved surfaces. This result is consistent with
previous findings on the homogeneity of asteroid families. For
example, it was previously found that asteroids within families can
share similar spectral properties \citep{CellinoA3} and colors
\citep{colors}. The tendency toward family homogeneity might be
helpful in deriving the family membership. Note that this tendency
does not support the claim that asteroid families originate from
differentiated parent bodies, since objects resulting from the
disruption of a differentiated parent body would show differing
photometric phase curves.  Therefore, the distribution of the $G_{12}$
values could contribute to the understanding the origin and evolution
of asteroid families.  $G_{12}$ could also be used, along with the
proper elements, for asteroid family classification. The trend from
smaller average $G_{12}$ values for the inner belt to larger $G_{12}$
values for the outer belt is consistent with the distribution of C and
S class asteroids in the asteroid belt. We also note that the $G_{12}$
values of family members in
Fig.~\ref{proper} 
correlate well with the SDSS color-color plot \citep{colors}. The
correlation relates to the fact that both the SDSS colors and $G_{12}$
correlate with asteroid taxonomy.

\begin{figure}[htbp]
   \centering
   \includegraphics[width=\textwidth]{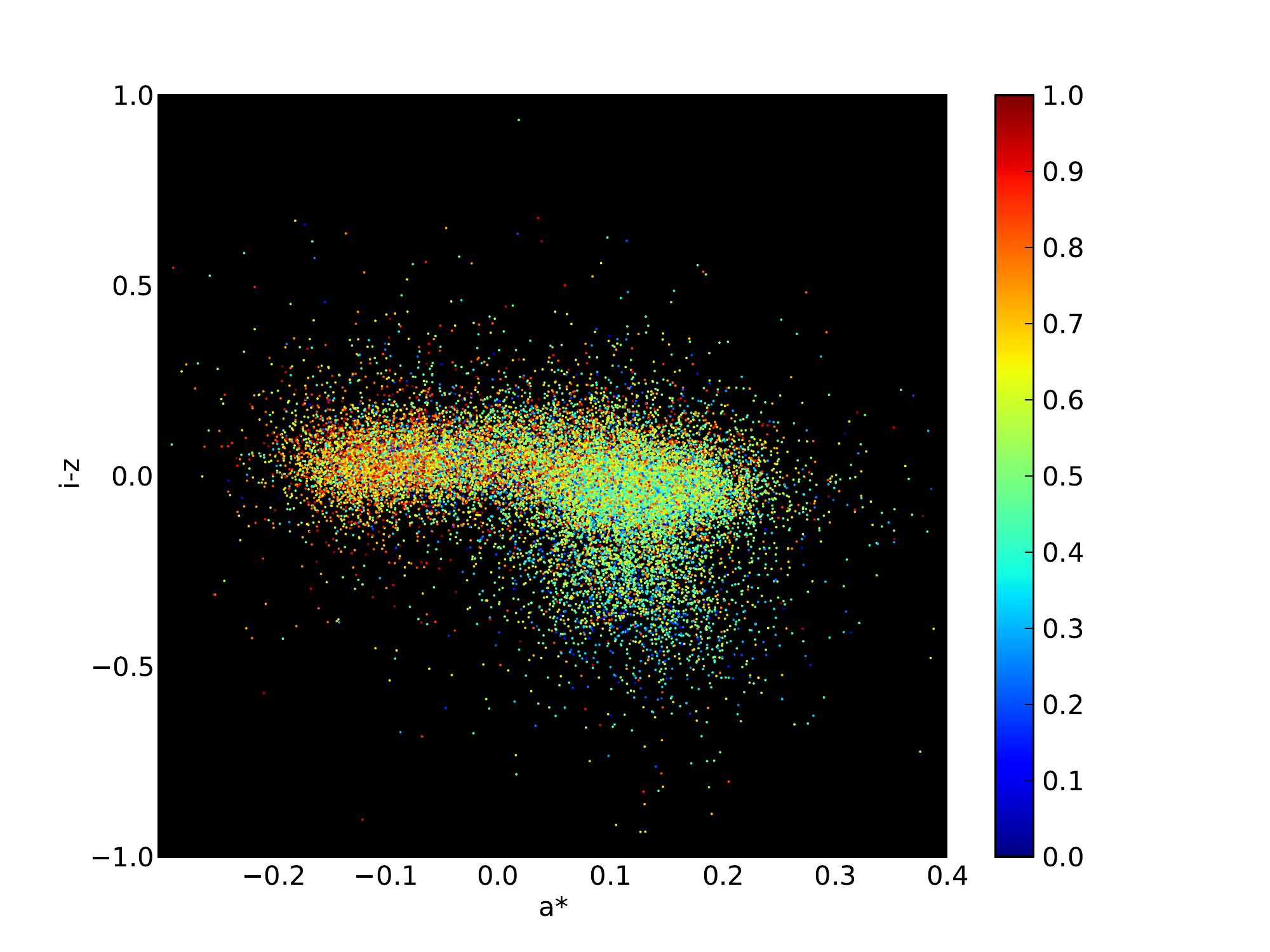} 
   \caption{Distribution of asteroids in $(a^*, i\!-\!z)$ SDSS color
     space, color-coded according to the $G_{12}$ value.}
   \label{SDSS}
\end{figure}

In Fig.~\ref{SDSS}, we plot the distribution of asteroids in SDSS
color-color space, coded according to the $G_{12}$ value. The $x$-axis
is defined as $a^* = 0.89 (g - r) + 0.45 (r - i) - 0.57$ and $y$-axis
as $i\!-\!z$, where $g$, $r$, $i$, and $z$ are magnitudes in the SDSS
filters. The two clouds correspond to the C and S class asteroids, and
the V class asteroids are located in the lower right corner of the
plot (with large $a^*$ and small $i\!-\!z$ values).  C class asteroids
tend to have, on average, larger values of $G_{12}$, S class smaller,
and V class often very small $G_{12}$ values.

To investigate the correlation of $G_{12}$ with taxonomy, we further
extracted taxonomic classifications from PDS.  The data set contains
entries for 2615 objects.  Each of the eight taxonomies represented
produced classifications for a subset of the objects: \cite{Tholena,
  Tholenb} -- 978 objects; \cite{Barucci} -- 438 objects;
\cite{Tedescoa}, \cite{Tedescob} -- 357 objects; \cite{Howell} -- 112
objects; \cite{Xu} -- 221 objects; \cite{Bus} -- 1447 objects;
\cite{Lazzaro} -- 820 objects; and \cite{DeMeo} -- 371 objects. We
make use of the \citeauthor{Bus} classification, which contains the
largest number of asteroids. We divide our sample into thirteen
complexes: A, C [B, C, Cb, Cg, Ch, Cgh], D [D, T], E [E, Xe], M, P, O,
Q [Q, Sq], R, S [S, Sa, Sk, Sl, Sr, K, L, Ld], V, X [X, Xc, Xk], and
U. We produce histograms of the $G_{12}$ values for each of them (see
Fig.~\ref{taxaHisto}). Each taxonomic complex is then approximated by
a Gaussian distribution. The means and standard deviations of the
$G_{12}$ values for all the complexes are listed in Table
\ref{G12taxa}. Most of the complexes contain too few objects for
meaningful statistical treatment, except for the S, C, and X
complexes. The means of the distributions for the S, C, and X
complexes are clearly different.

The S complex has a mean $G_{12}$ of $0.41$, the C complex has a
higher mean $G_{12}$ of $0.64$, and the X complex is intermediate
having a mean of $0.48$. In general, asteroids within the same
taxonomic complex could have varying surface properties (for example,
different regolith porosities or grain-size distributions) leading to
different $G_{12}$ values, resulting in broad $G_{12}$ histograms for
a complex. An additional challenge follows from the fact that the
$G_{12}$ distributions for the different taxonomic complexes partially
overlap. Based on those distributions, selected priors (see
Sec.~\ref{method}) and previously obtained photometric parameters
\citep{DO} for each of the half a million asteroids, we computed the
C, S, and X complex classification probabilities for each
asteroid. Due to broad and overlapping $G_{12}$ distributions, these
probabilities are often be similar enough to prevent a meaningful
classification of the asteroid into any one of the complexes.

For some asteroids, $G_{12}$ can, however, be a good indicator of
taxonomic complex. For example, an asteroid with $G_{12} = 0.8$ from
the outer belt has a probability of 82\% for being of C complex and
low probabilities of being of S or X complex (5\% and 13\%).  If we
assume no knowledge on asteroid location nor on the frequency of
different taxonomic complexes (uniform prior (1)), an asteroid with
$G_{12} = 0.8$ would still have a chance of 70\% for being of C
complex. For reference, we list the probabilities for an asteroid with
$G_{12} = 0.8$ being of C, S, and X complex in Table~\ref{probEx},
assuming different priors and different locations in the belt (or no
knowledge on location in the belt).

\begin{table}[htb]
\centering
\caption{Example result for a single asteroid. Probabilities for an
  asteroid with $G_{12}=0.8$ to be of C, S, and X complex.}
\label{probEx}
\hspace{1ex}
\begin{tabular}{llll}
Prior & $P_c$ & $P_s$ & $P_x$ \\ \hline
(1) & 70\% & 6\% & 24\% \\
(2) & 76\% & 10\% & 14\% \\
(3) Inner MB & 66\% & 21\% & 13\% \\
(3) Mid MB & 79\% & 7\% & 14\% \\
(3) Outer MB & 82\% & 5\% & 13\% \\ \hline
\end{tabular}
\end{table}

Asteroid families containing, for example, a large number of asteroids
with high $G_{12}$ values resulting in high $P_c$ values could be
identified as C-complex preponderant. As a prime example, we indicate
the Dora family having the mean $G_{12}=0.7$ and standard deviation
$\sigma_{G_{12}}=0.18$, which result in a high C-complex preponderance
probability. The $G_{12}$ distribution (Fig.~\ref{family}) for the
Dora family also matches nicely the C-complex distribution profile.

\begin{center}
   \begin{landscape}
        \begin{longtable}{lllllllllllll}
          \caption{The number of family members included in the study,
            the mean and standard deviation of $G_{12}$, and the
            probabilities of C, S, and X complex preponderance based
            on the a priori distributions (1), (2), and (3) (see
            above). Family classifications are from \cite{Nesvorny}.}
          \label{mean} \\
	& \multicolumn{3}{c}{General} & \multicolumn{3}{c}{(1) Uniform} & \multicolumn{3}{c}{(2) Frequency} & \multicolumn{3}{c}{(3) Location} \\
	& \multicolumn{3}{c}{$G_{12}$ statistics} & \multicolumn{3}{c}{prior} & \multicolumn{3}{c}{prior} & \multicolumn{3}{c}{prior}  \\
	Family or & Nr of & $G_{12}$ & $G_{12}$ & $P_{C}$ & $P_{S}$ & $P_{X}$ & $P_{C}$ & $P_{S}$ & $P_{X}$ & $P_{C}$ & $P_{S}$ & $P_{X}$ \\ 
	 	   cluster &  mem. & mean & std. & & & & & & & &  \\ \hline
Adeona & 987 & 0.64 & 0.2 & 0.49 & 0.21 & 0.29 & 0.55 & 0.27 & 0.18 & 0.55 & 0.27 & 0.18 \\ 
Aeolia & 55 & 0.66 & 0.23 & 0.51 & 0.2 & 0.29 & 0.54 & 0.29 & 0.16 & 0.57 & 0.25 & 0.17 \\ 
Agnia & 472 & 0.58 & 0.21 & 0.42 & 0.27 & 0.31 & 0.47 & 0.34 & 0.19 & 0.47 & 0.34 & 0.19 \\ 
Astrid & 94 & 0.64 & 0.23 & 0.49 & 0.22 & 0.29 & 0.43 & 0.45 & 0.12 & 0.55 & 0.27 & 0.17 \\ 
Baptistina & 1966 & 0.53 & 0.18 & 0.36 & 0.31 & 0.33 & 0.46 & 0.32 & 0.22 & 0.26 & 0.62 & 0.12 \\ 
Beagle & 38 & 0.64 & 0.23 & 0.5 & 0.21 & 0.29 & 0.6 & 0.21 & 0.18 & 0.6 & 0.21 & 0.18 \\ 
Brang\"{a}ne & 37 & 0.68 & 0.19 & 0.53 & 0.19 & 0.28 & 0.64 & 0.19 & 0.18 & 0.6 & 0.24 & 0.17 \\ 
Brasilia & 186 & 0.49 & 0.21 & 0.31 & 0.35 & 0.34 & 0.4 & 0.37 & 0.23 & 0.4 & 0.37 & 0.23 \\ 
Charis & 136 & 0.61 & 0.21 & 0.45 & 0.25 & 0.31 & 0.55 & 0.25 & 0.2 & 0.55 & 0.25 & 0.2   \\ 
Chloris & 200 & 0.63 & 0.17 & 0.48 & 0.22 & 0.3 & 0.59 & 0.22 & 0.19 & 0.55 & 0.28 & 0.18 \\ 
Clarissa & 41 & 0.64 & 0.22 & 0.49 & 0.22 & 0.29 & 0.42 & 0.46 & 0.12 & 0.42 & 0.46 & 0.12  \\ 
Datura & 4 & 0.41 & 0.2 & 0.23 & 0.42 & 0.35 & 0.3 & 0.45 & 0.25 & 0.17 & 0.72 & 0.11 \\ 
Dora & 528 & 0.7 & 0.18 & 0.56 & 0.16 & 0.28 & 0.67 & 0.16 & 0.17 & 0.63 & 0.2 & 0.16 \\ 
Emma & 111 & 0.66 & 0.23 & 0.52 & 0.19 & 0.29 & 0.59 & 0.24 & 0.17 & 0.63 & 0.19 & 0.18   \\
Emilkowalski & 2 & 0.56 & 0.07 & 0.39 & 0.28 & 0.33 & 0.45 & 0.35 & 0.2 & 0.45 & 0.35 & 0.2   \\ 
Eos & 3413 & 0.64 & 0.19 & 0.5 & 0.21 & 0.29 & 0.56 & 0.27 & 0.18 & 0.6 & 0.21 & 0.19 \\ 
Erigone & 806 & 0.63 & 0.19 & 0.48 & 0.22 & 0.3 & 0.54 & 0.28 & 0.18 & 0.4 & 0.48 & 0.12  \\ 
Eunomia & 4707 & 0.51 & 0.18 & 0.33 & 0.34 & 0.33 & 0.37 & 0.42 & 0.2 & 0.37 & 0.42 & 0.2 \\ 
Flora & 6316 & 0.53 & 0.18 & 0.35 & 0.32 & 0.33 & 0.26 & 0.63 & 0.11 & 0.26 & 0.63 & 0.11 \\ 
Gefion & 1938 & 0.56 & 0.19 & 0.39 & 0.29 & 0.32 & 0.49 & 0.3 & 0.21 & 0.44 & 0.36 & 0.19 \\ 
Hestia & 103 & 0.55 & 0.2 & 0.39 & 0.29 & 0.32 & 0.44 & 0.37 & 0.19 & 0.44 & 0.37 & 0.19  \\ 
Hoffmeister & 341 & 0.68 & 0.22 & 0.53 & 0.19 & 0.28 & 0.63 & 0.19 & 0.18 & 0.59 & 0.24 & 0.17 \\ 
Hygiea & 1729 & 0.66 & 0.21 & 0.51 & 0.2 & 0.29 & 0.61 & 0.2 & 0.18 & 0.61 & 0.2 & 0.18 \\ 
Iannini & 30 & 0.45 & 0.23 & 0.26 & 0.4 & 0.34 & 0.29 & 0.5 & 0.21 & 0.29 & 0.5 & 0.21 \\ 
Juno & 359 & 0.52 & 0.21 & 0.36 & 0.32 & 0.33 & 0.27 & 0.61 & 0.11 & 0.4 & 0.4 & 0.2   \\ 
Karin & 159 & 0.54 & 0.22 & 0.38 & 0.3 & 0.32 & 0.29 & 0.59 & 0.12 & 0.47 & 0.31 & 0.22   \\ 
Kazvia & 12 & 0.6 & 0.21 & 0.45 & 0.24 & 0.31 & 0.51 & 0.31 & 0.18 & 0.51 & 0.31 & 0.18   \\ 
Konig & 58 & 0.68 & 0.17 & 0.53 & 0.18 & 0.28 & 0.6 & 0.23 & 0.17 & 0.6 & 0.23 & 0.17 \\ 
Koronis & 2913 & 0.58 & 0.2 & 0.42 & 0.27 & 0.31 & 0.52 & 0.28 & 0.21 & 0.52 & 0.28 & 0.21 \\ 
Lau & 6 & 0.54 & 0.17 & 0.39 & 0.28 & 0.33 & 0.5 & 0.28 & 0.21 & 0.5 & 0.28 & 0.21 \\ 
Lixiaohua & 171 & 0.64 & 0.23 & 0.5 & 0.21 & 0.29 & 0.43 & 0.45 & 0.12 & 0.59 & 0.22 & 0.19  \\ 
Lucienne & 37 & 0.5 & 0.21 & 0.33 & 0.34 & 0.33 & 0.25 & 0.64 & 0.11 & 0.25 & 0.64 & 0.11  \\ 
Maria & 2009 & 0.5 & 0.2 & 0.33 & 0.34 & 0.33 & 0.37 & 0.42 & 0.2 & 0.37 & 0.42 & 0.2  \\ 
Massalia & 1911 & 0.56 & 0.2 & 0.39 & 0.29 & 0.32 & 0.44 & 0.37 & 0.19 & 0.3 & 0.58 & 0.12   \\ 
Meliboea & 40 & 0.68 & 0.14 & 0.55 & 0.17 & 0.28 & 0.62 & 0.21 & 0.17 & 0.67 & 0.16 & 0.17\\ 
Merxia & 425 & 0.53 & 0.22 & 0.36 & 0.31 & 0.33 & 0.46 & 0.33 & 0.22 & 0.41 & 0.39 & 0.2 \\ 
Misa & 220 & 0.68 & 0.21 & 0.54 & 0.18 & 0.28 & 0.6 & 0.23 & 0.17 & 0.6 & 0.23 & 0.17 \\ 
Naema & 98 & 0.66 & 0.2 & 0.51 & 0.2 & 0.29 & 0.58 & 0.25 & 0.17 & 0.62 & 0.2 & 0.18 \\ 
Nemesis & 258 & 0.68 & 0.19 & 0.54 & 0.18 & 0.28 & 0.64 & 0.18 & 0.18 & 0.6 & 0.23 & 0.17 \\ 
Nysa-Polana & 8289 & 0.57 & 0.19 & 0.4 & 0.28 & 0.32 & 0.46 & 0.35 & 0.19 & 0.31 & 0.57 & 0.12 \\ 
Padua & 372 & 0.66 & 0.2 & 0.52 & 0.19 & 0.29 & 0.59 & 0.24 & 0.17 & 0.59 & 0.24 & 0.17 \\ 
Rafita & 477 & 0.55 & 0.2 & 0.39 & 0.29 & 0.32 & 0.48 & 0.3 & 0.21 & 0.44 & 0.37 & 0.19 \\ 
Sulamitis & 92 & 0.66 & 0.21 & 0.52 & 0.2 & 0.29 & 0.62 & 0.2 & 0.18 & 0.45 & 0.43 & 0.12 \\ 
Sylvia & 30 & 0.56 & 0.26 & 0.43 & 0.25 & 0.31 & 0.53 & 0.27 & 0.21 & 0.46 & 0.37 & 0.17 \\ 
Telramund & 240 & 0.58 & 0.23 & 0.42 & 0.27 & 0.31 & 0.47 & 0.34 & 0.19 & 0.51 & 0.28 & 0.21 \\ 
Terentia & 7 & 0.41 & 0.2 & 0.24 & 0.4 & 0.36 & 0.15 & 0.74 & 0.11 & 0.32 & 0.42 & 0.25 \\ 
Themis & 2559 & 0.69 & 0.19 & 0.55 & 0.17 & 0.28 & 0.62 & 0.22 & 0.17 & 0.66 & 0.17 & 0.17  \\ 
Theobalda & 60 & 0.61 & 0.22 & 0.47 & 0.23 & 0.3 & 0.39 & 0.49 & 0.12 & 0.57 & 0.24 & 0.19   \\ 
Tirela & 818 & 0.61 & 0.2 & 0.46 & 0.24 & 0.3 & 0.56 & 0.24 & 0.2 & 0.56 & 0.24 & 0.2   \\ 
Veritas & 291 & 0.68 & 0.21 & 0.54 & 0.18 & 0.28 & 0.61 & 0.22 & 0.17 & 0.65 & 0.18 & 0.17 \\ 
Vesta & 8445 & 0.5 & 0.18 & 0.32 & 0.35 & 0.34 & 0.36 & 0.44 & 0.2 & 0.22 & 0.66 & 0.11 \\ 
18405 & 24 & 0.63 & 0.16 & 0.48 & 0.22 & 0.3 & 0.51 & 0.32 & 0.17 & 0.6 & 0.21 & 0.19  \\ 
18466 & 94 & 0.49 & 0.22 & 0.31 & 0.35 & 0.34 & 0.24 & 0.65 & 0.11 & 0.36 & 0.44 & 0.2  \\ 
\hline
   \end{longtable}
  \end{landscape}
\end{center}

In order to check how well we can identify asteroid families as being
dominated by one of the taxonomic complexes, we produce $G_{12}$
histograms for the different asteroid families (histograms for chosen families are presented in 
Fig.~\ref{family}, histograms for the remaining families can be found in 
supplementary materials; numerical values are in Table \ref{mean}) and use methods based on Bayesian
statistics (described in Sec.~\ref{methods}) to establish the dominant
taxonomic complex. We then compare our results with the published
results from other studies.

\begin{center}
\begin{longtable}{cc}
\hline
\includegraphics[width=0.5 \textwidth]{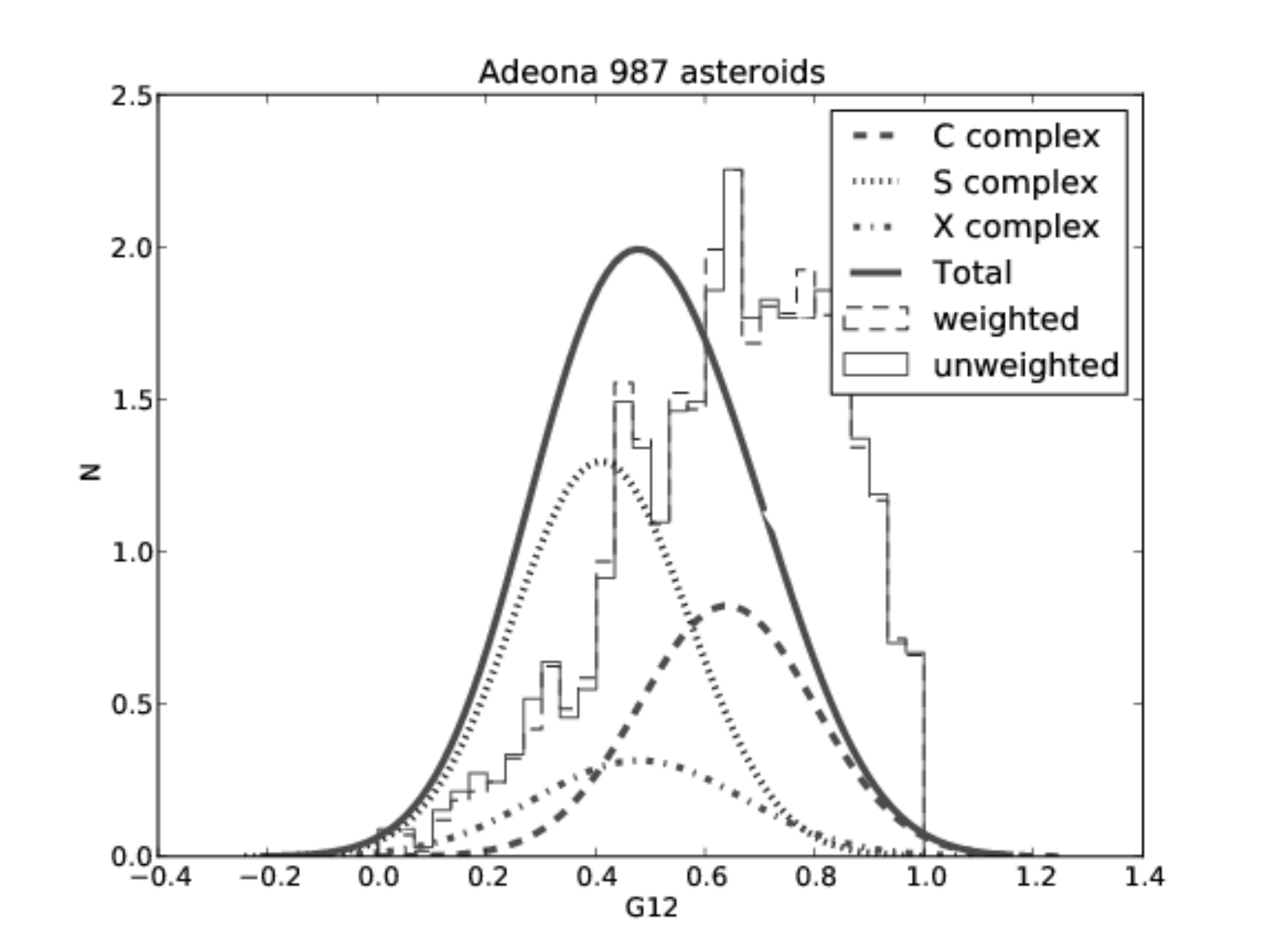} &
\includegraphics[width=0.5 \textwidth]{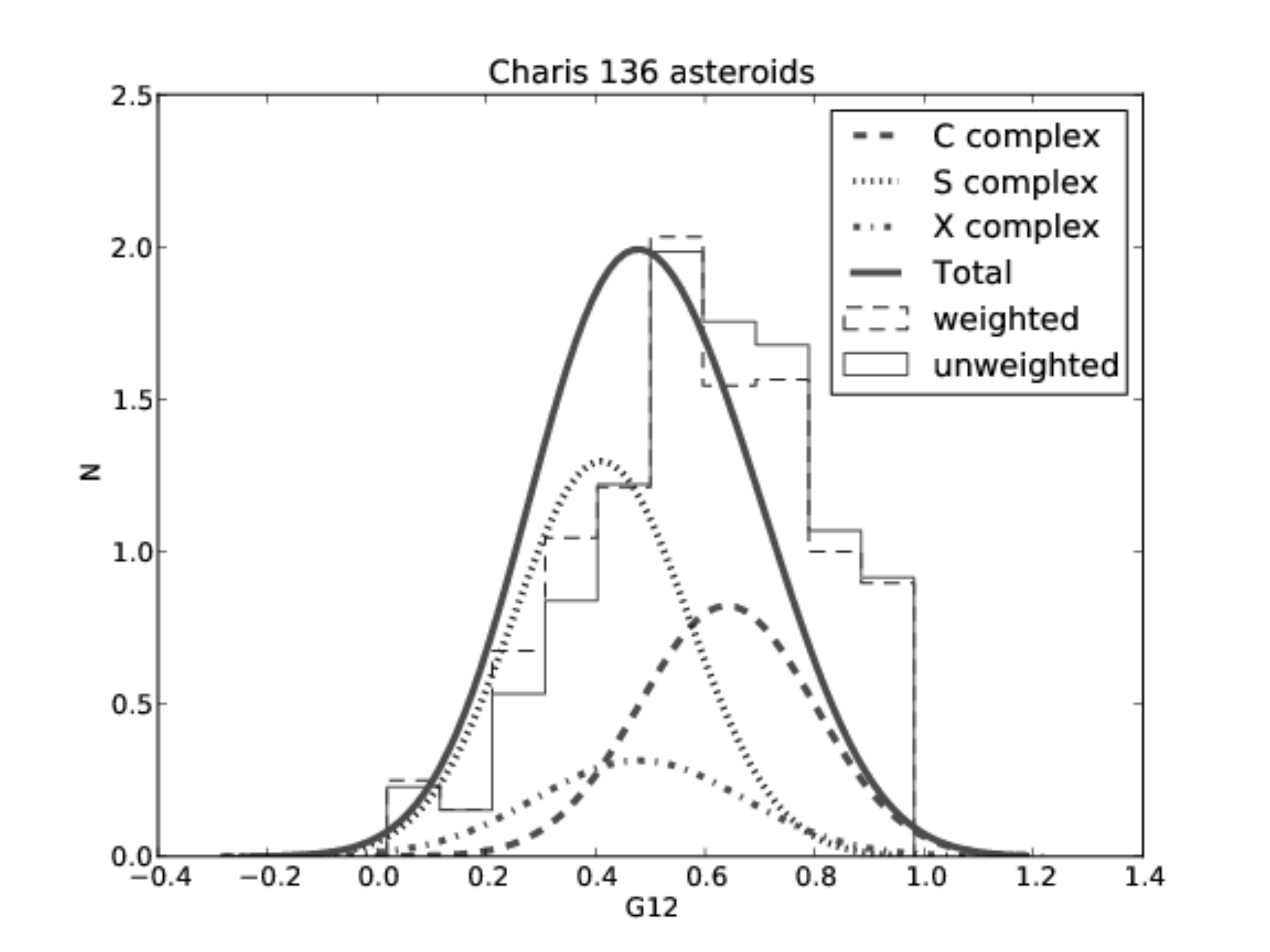} \\
\includegraphics[width=0.5 \textwidth]{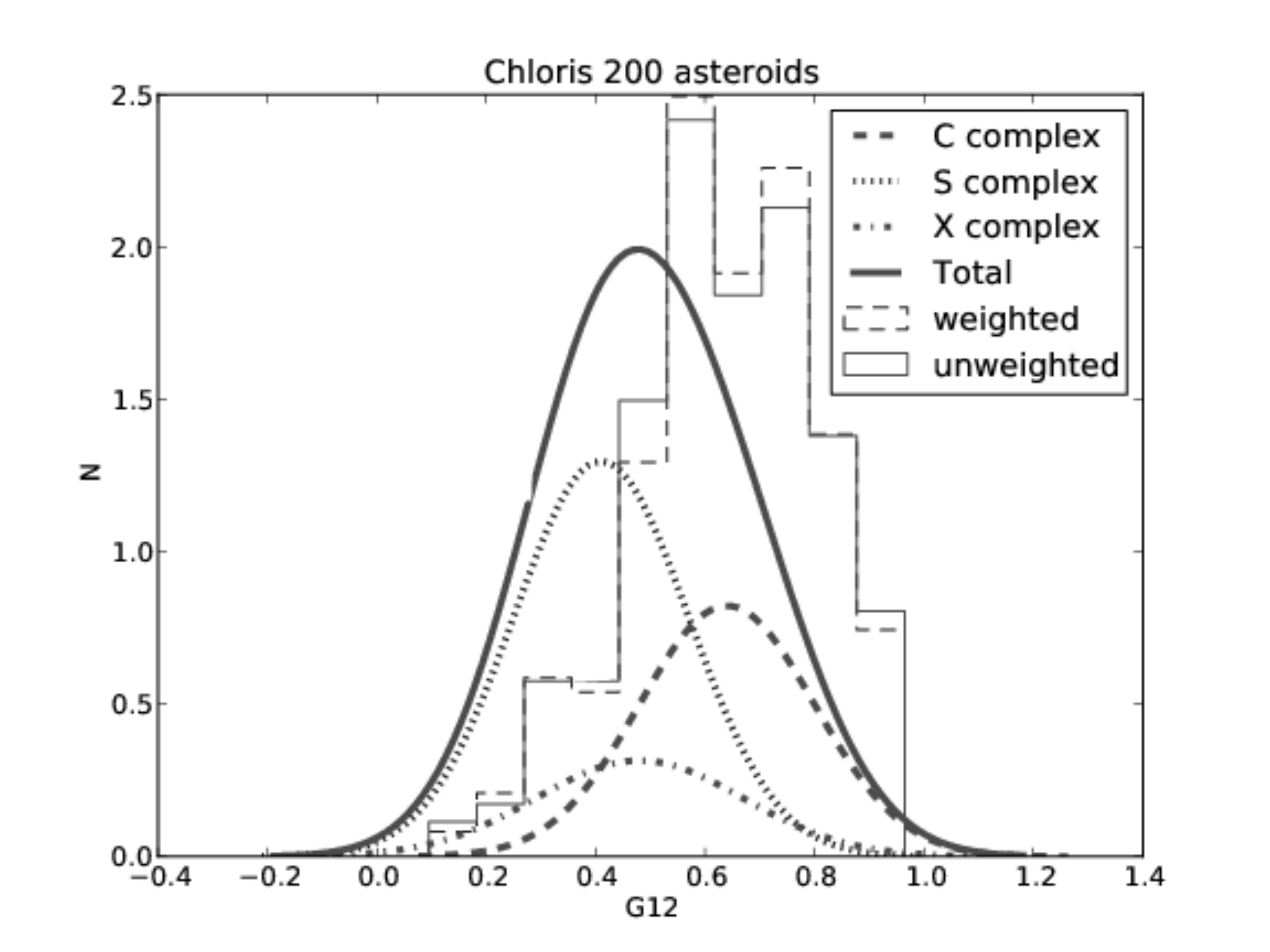} &
\includegraphics[width=0.5 \textwidth]{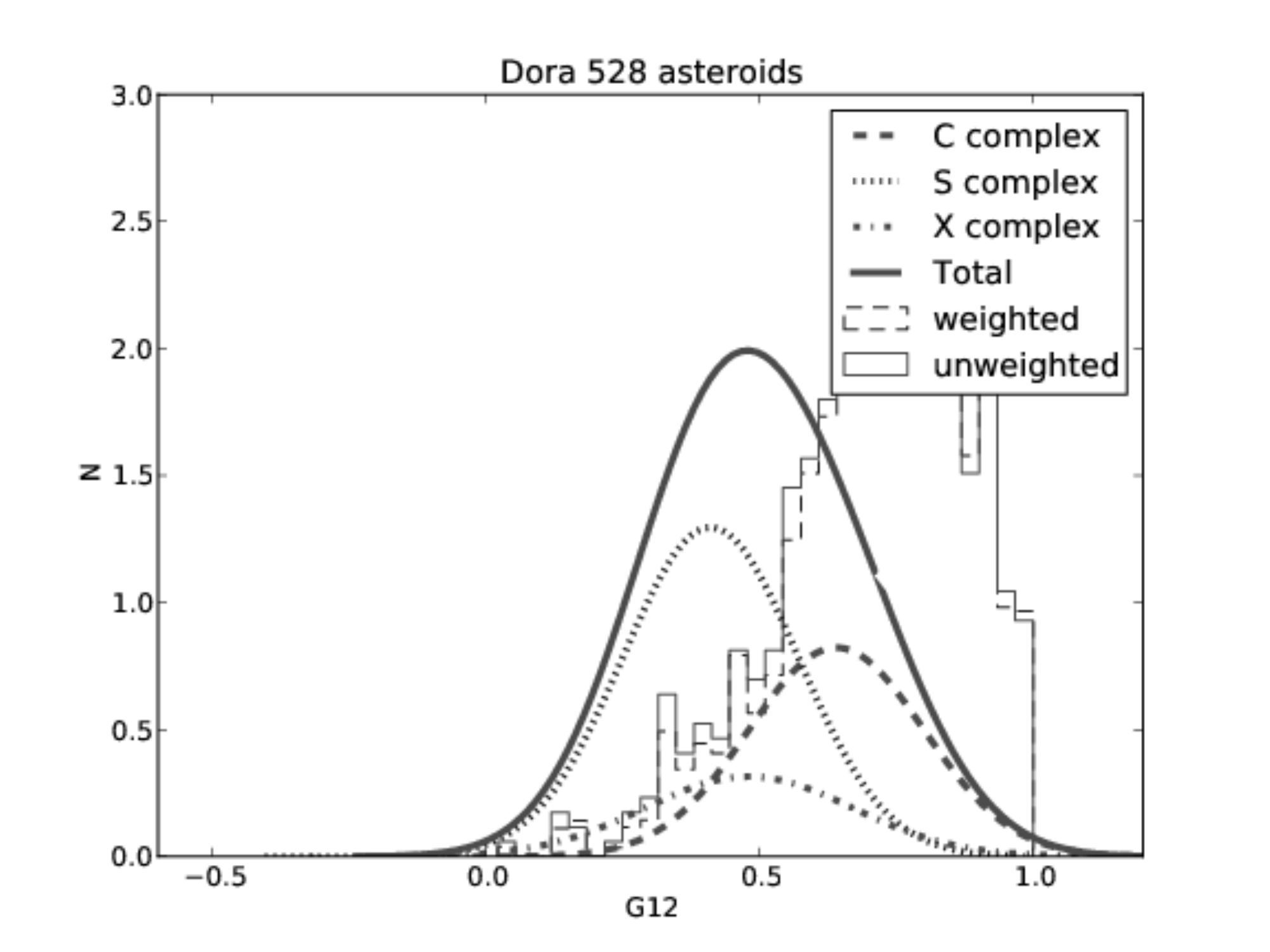} \\
\includegraphics[width=0.5 \textwidth]{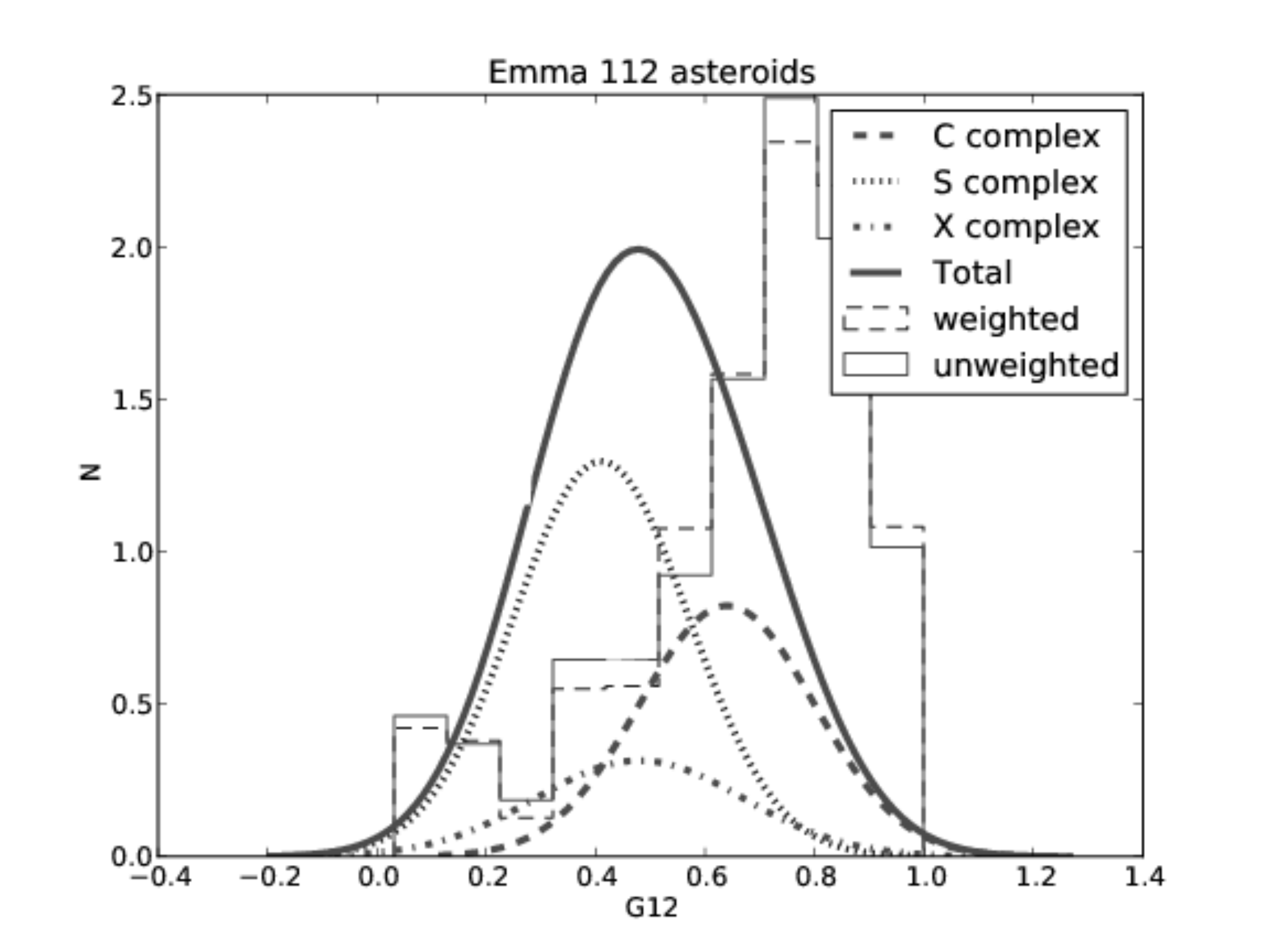} &
\includegraphics[width=0.5 \textwidth]{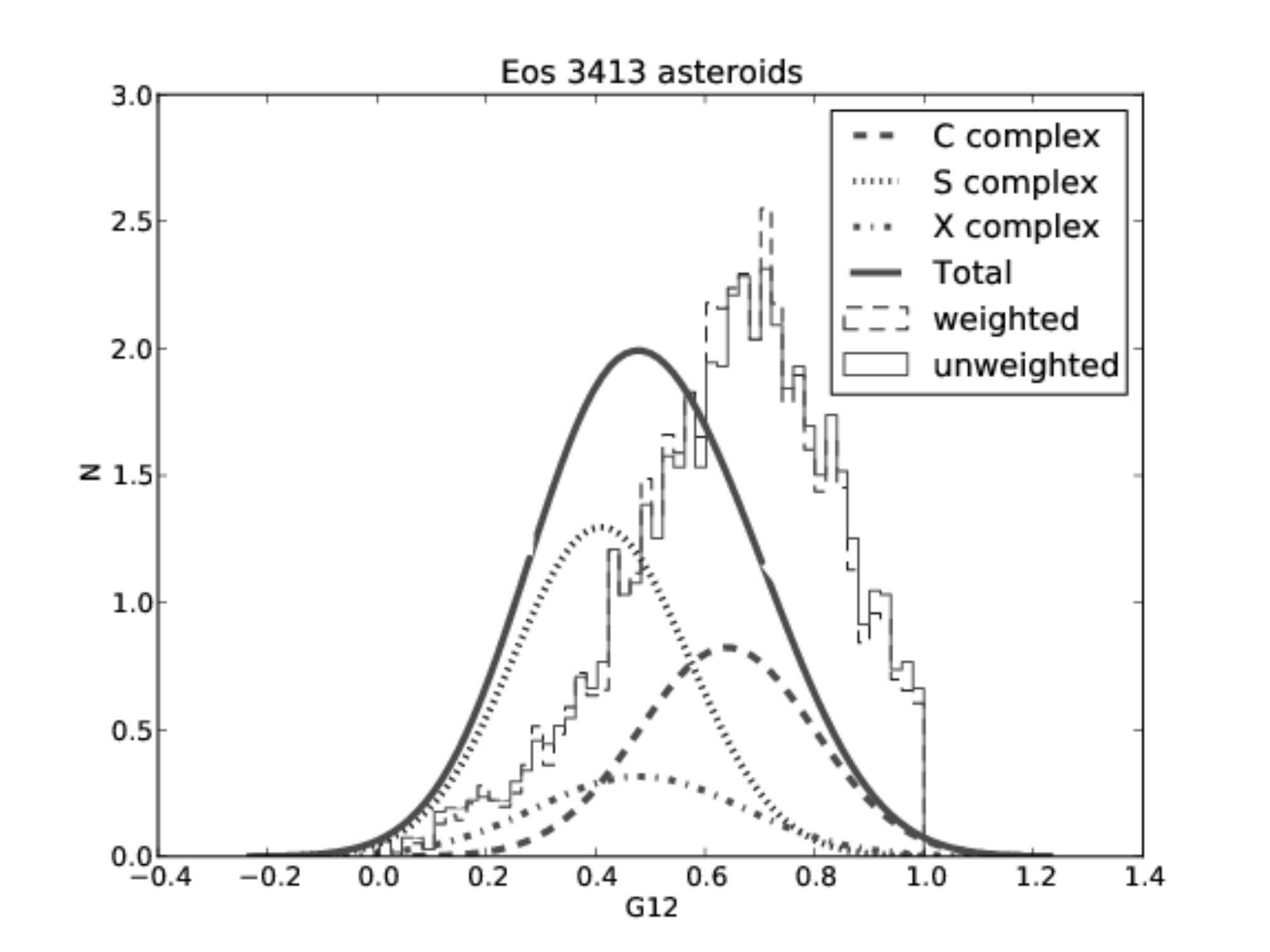}\\  
\includegraphics[width=0.5 \textwidth]{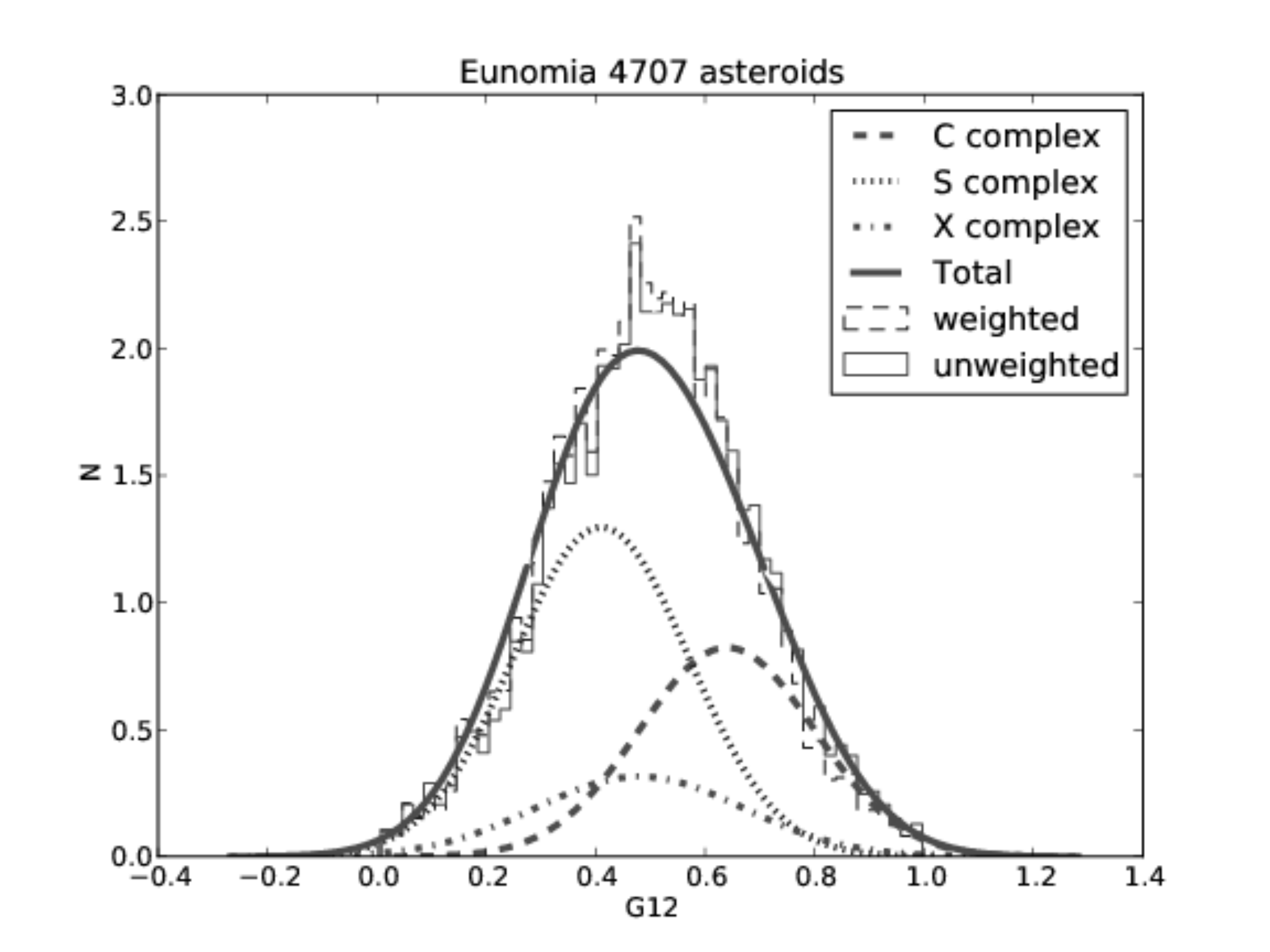} &
\includegraphics[width=0.5 \textwidth]{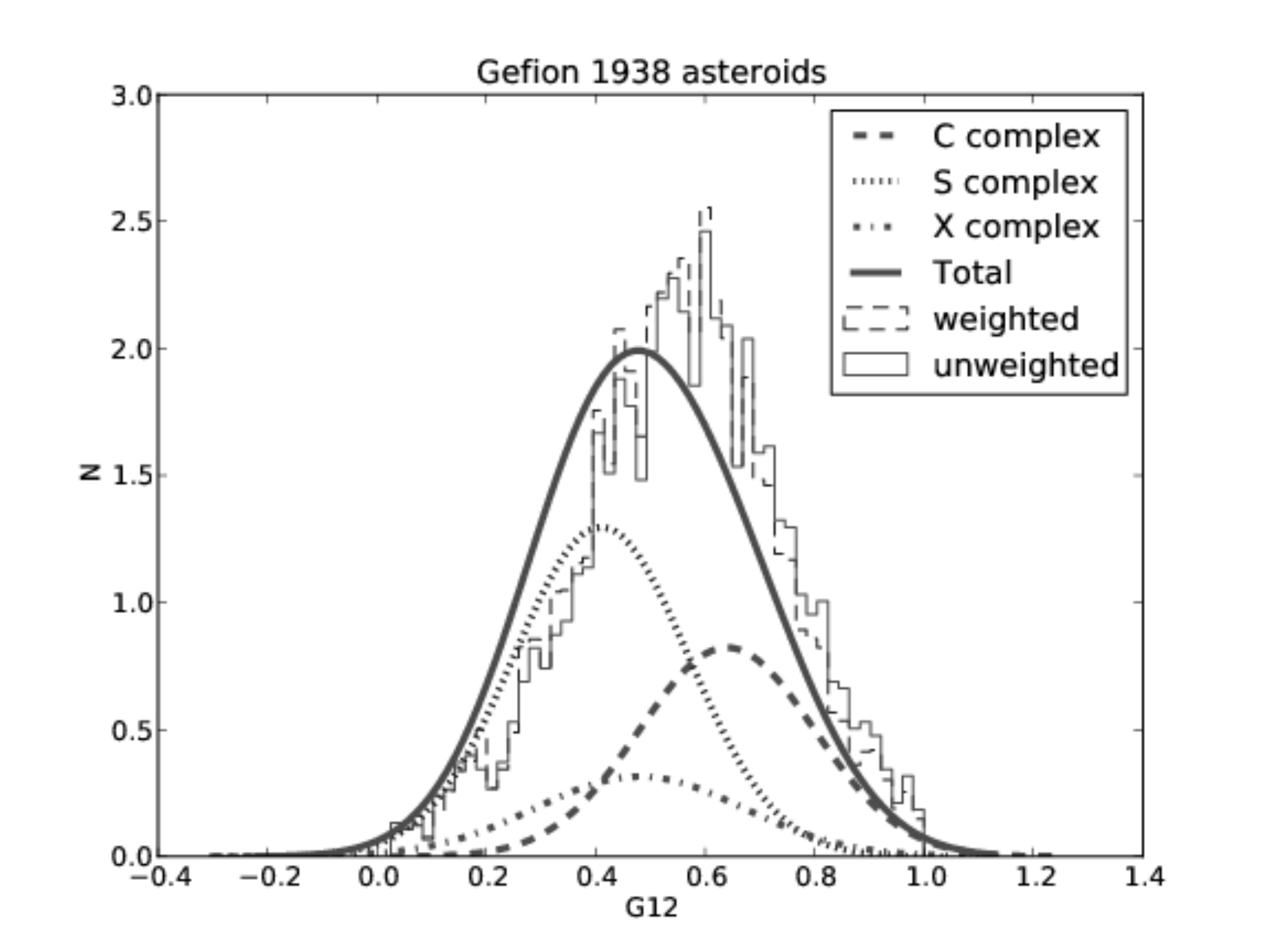} \\
\includegraphics[width=0.5 \textwidth]{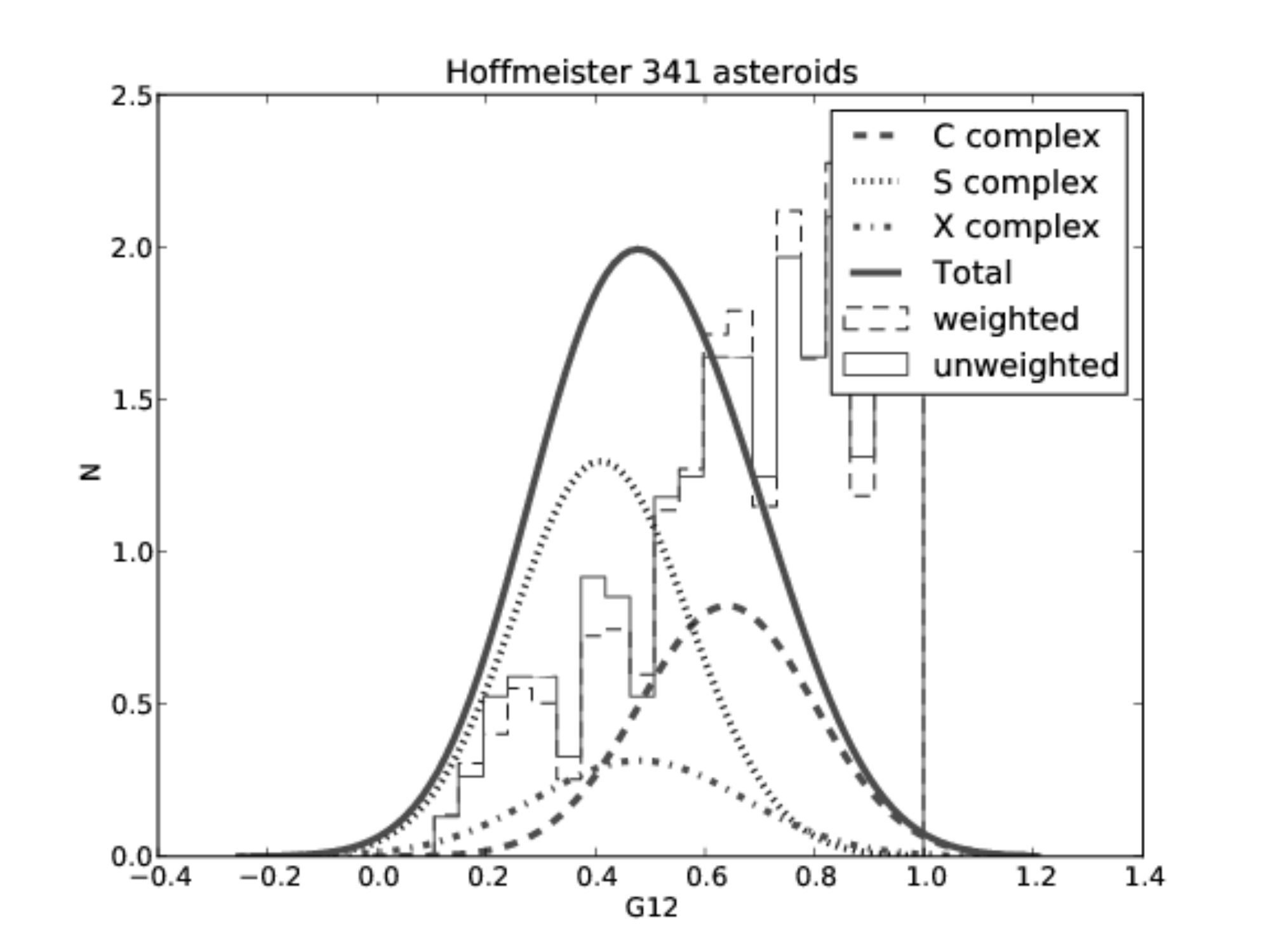} & 
\includegraphics[width=0.5 \textwidth]{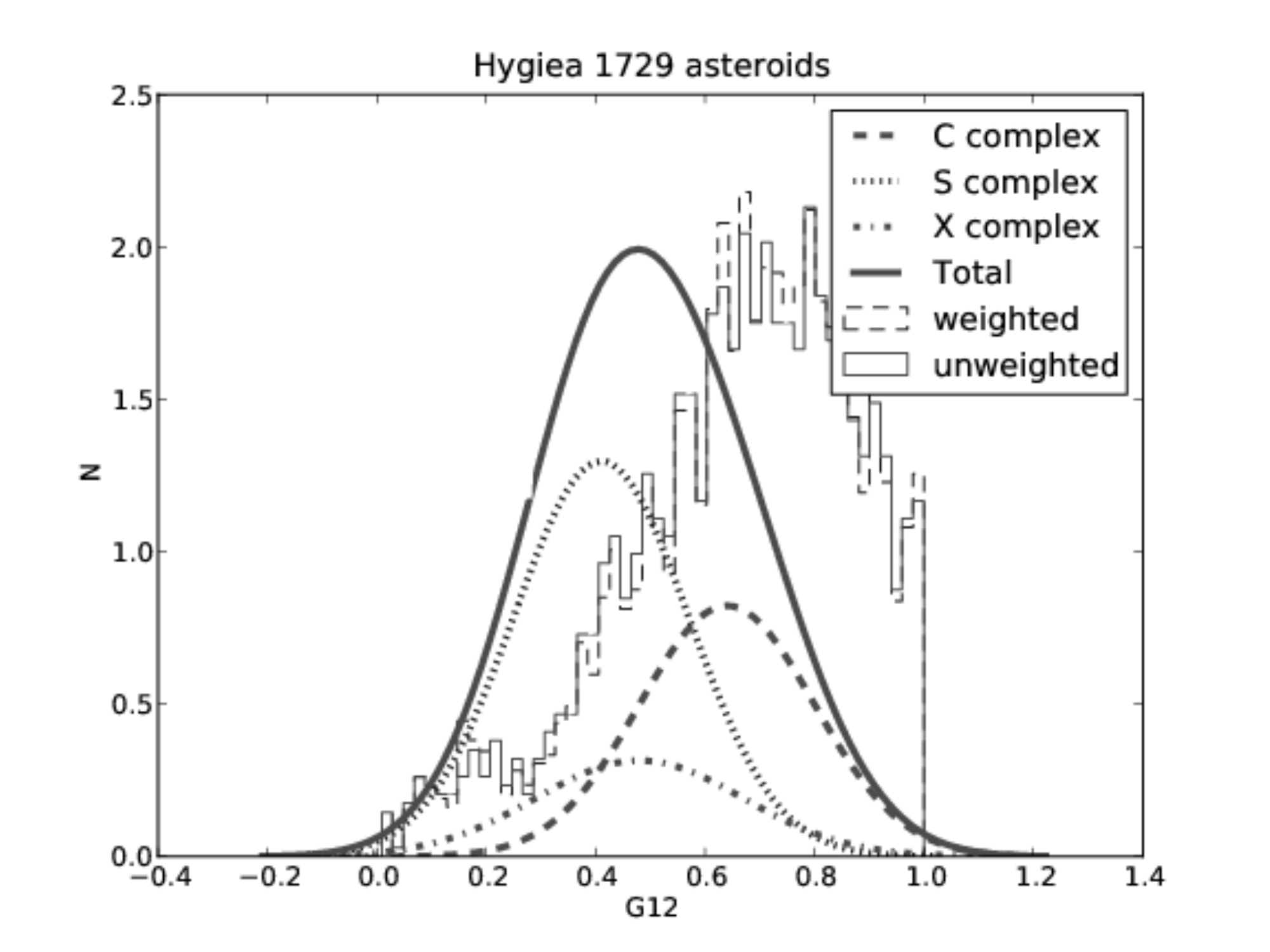} \\
\includegraphics[width=0.5 \textwidth]{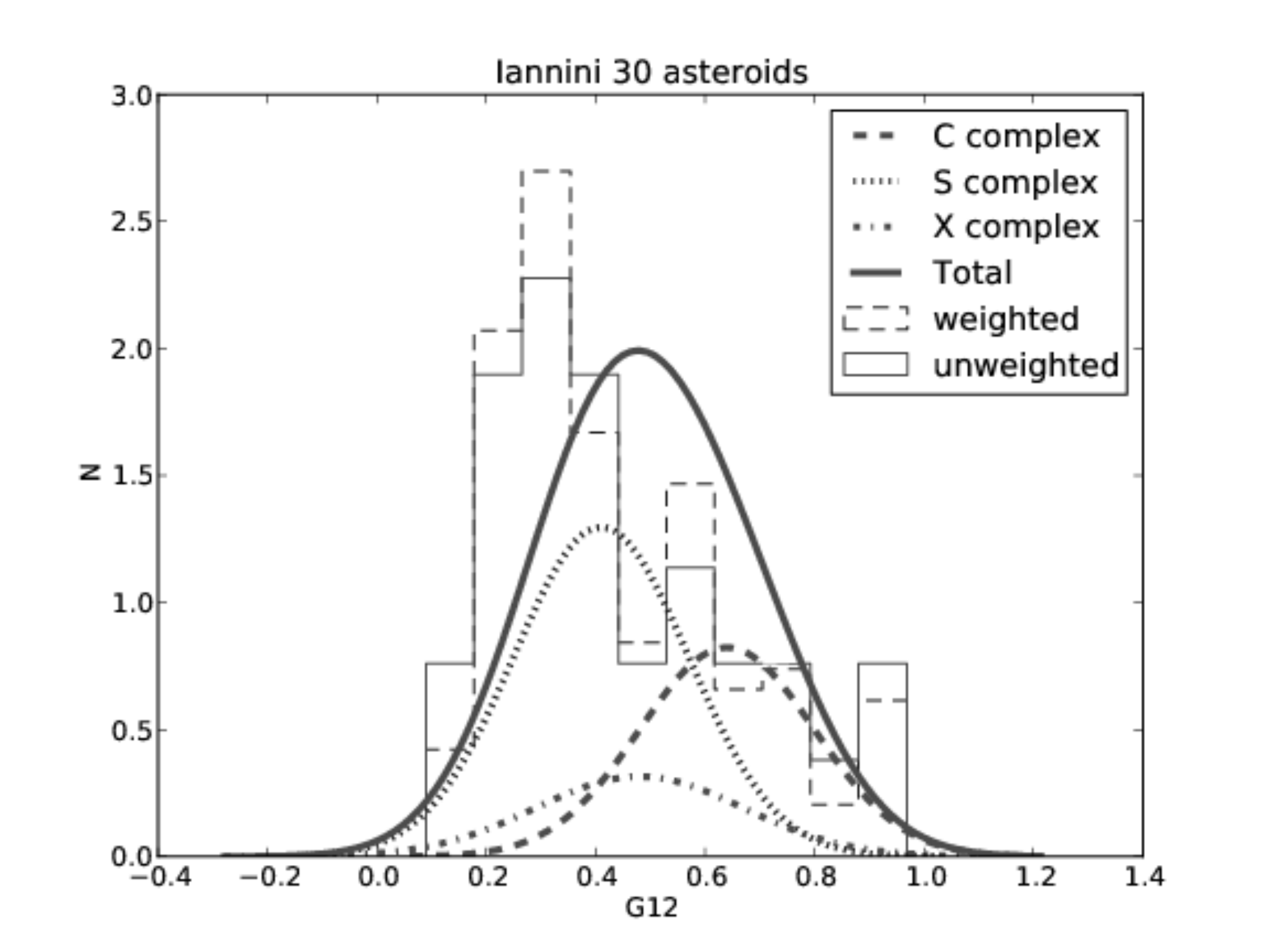} &
\includegraphics[width=0.5 \textwidth]{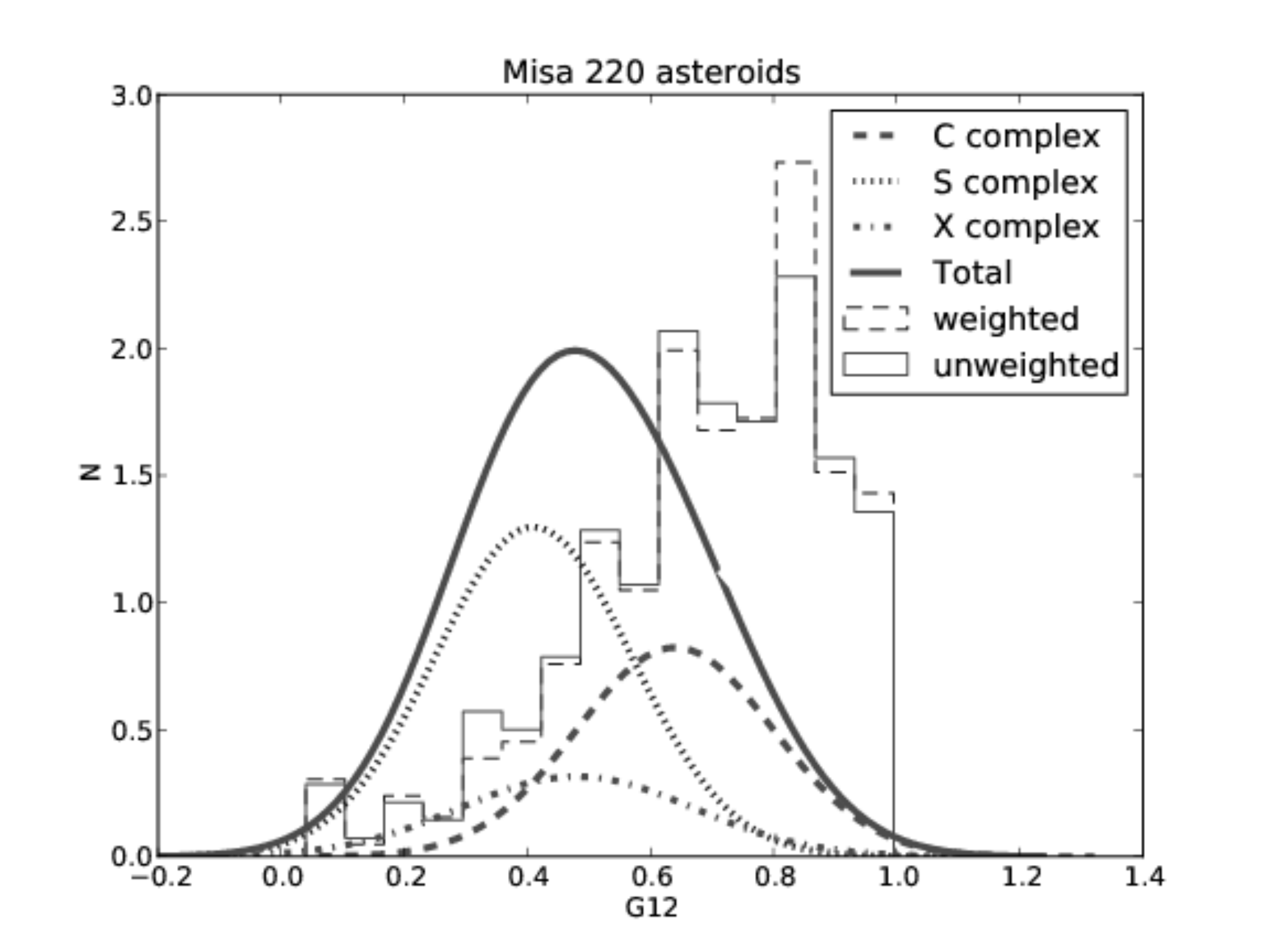} \\
\includegraphics[width=0.5 \textwidth]{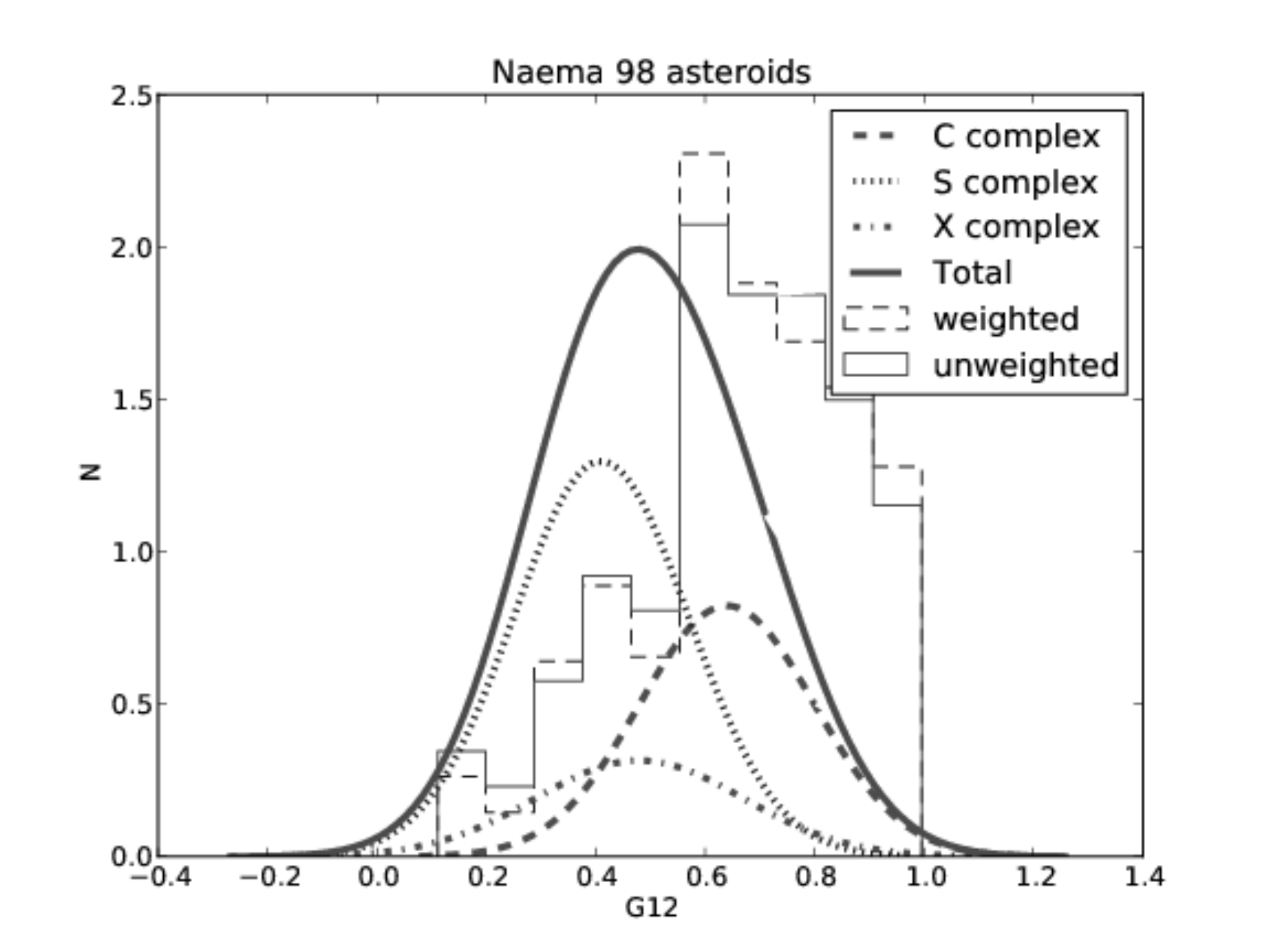} &
\includegraphics[width=0.5 \textwidth]{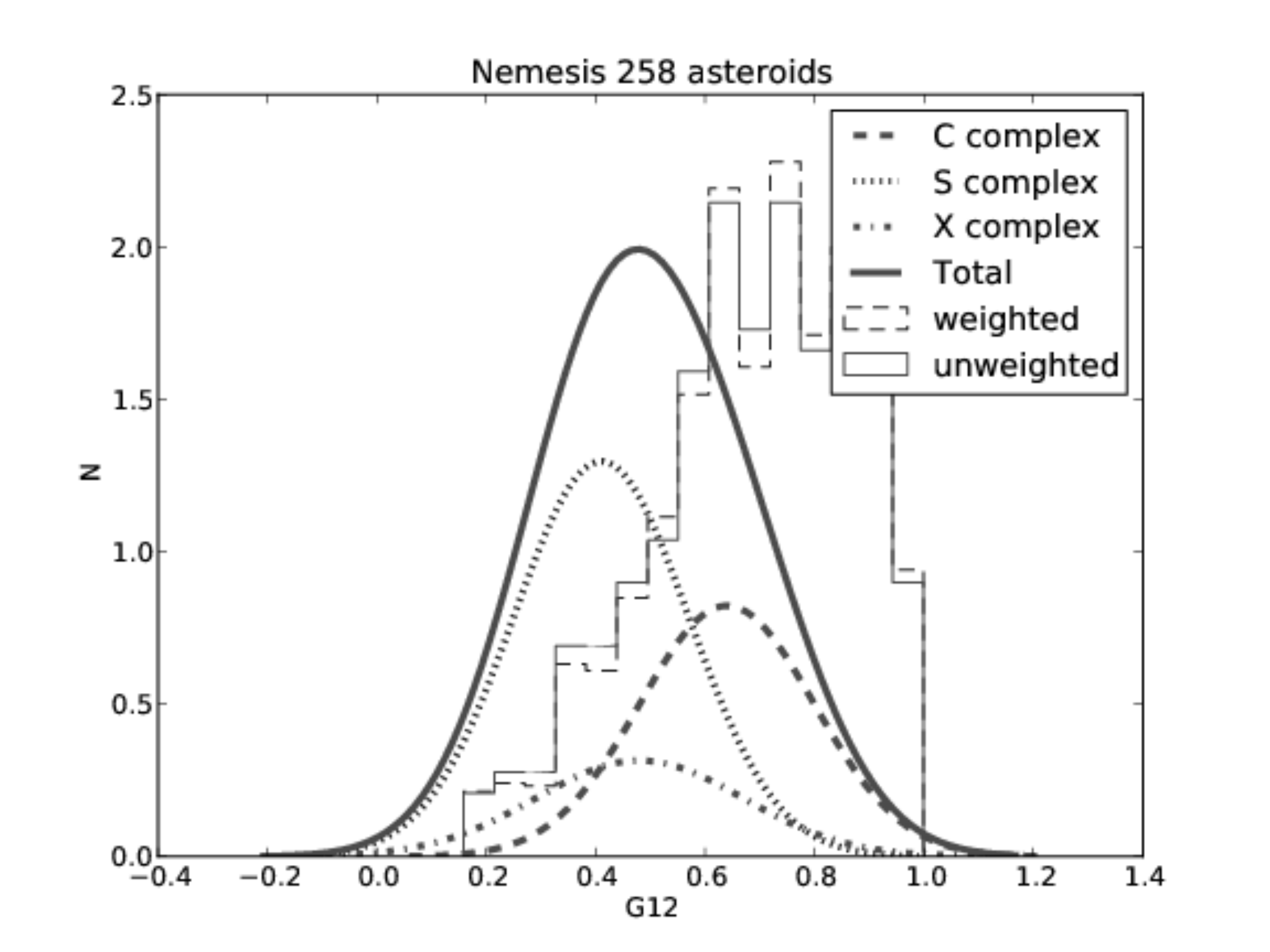} \\  
\includegraphics[width=0.5 \textwidth]{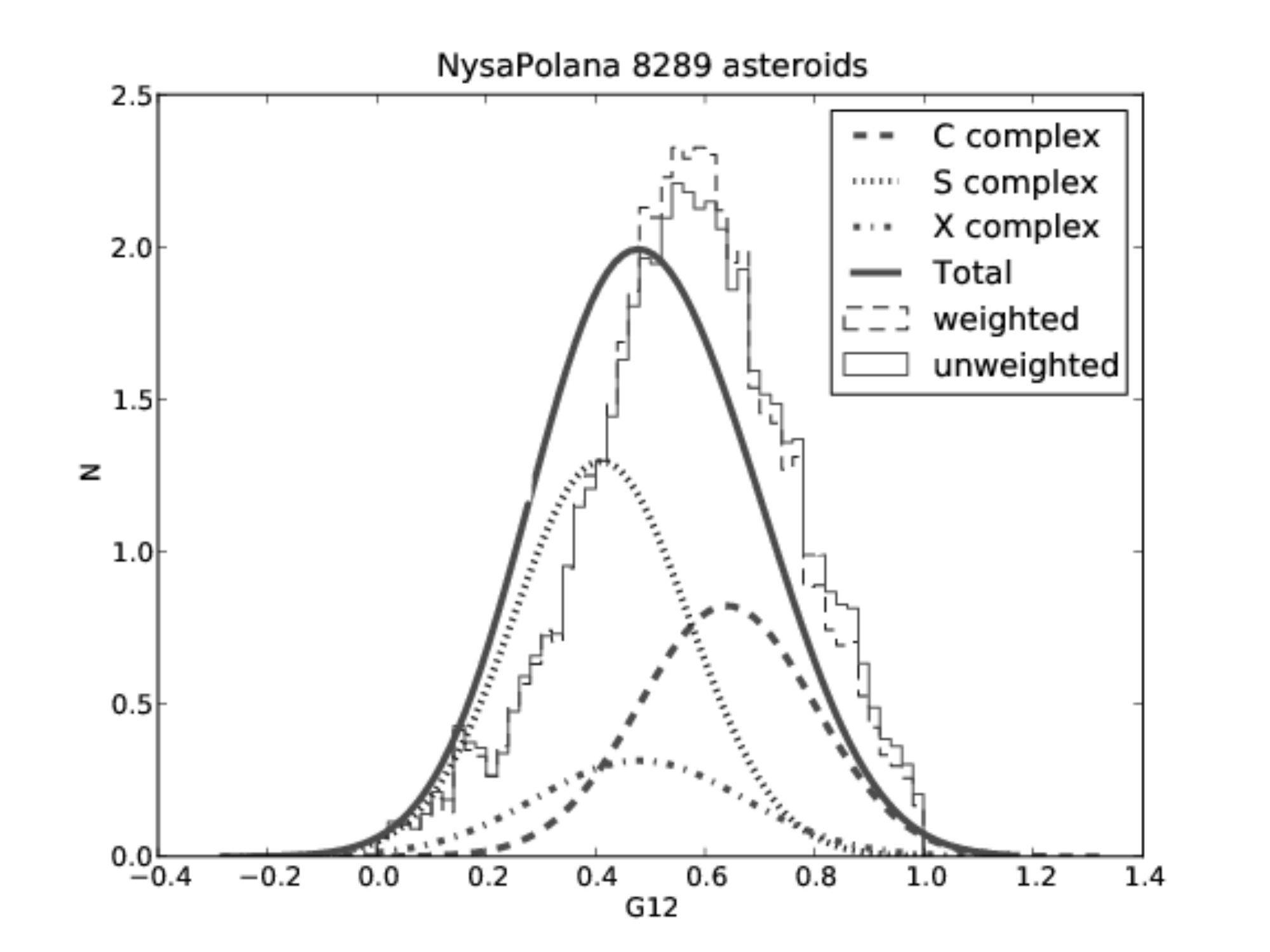} &
\includegraphics[width=0.5 \textwidth]{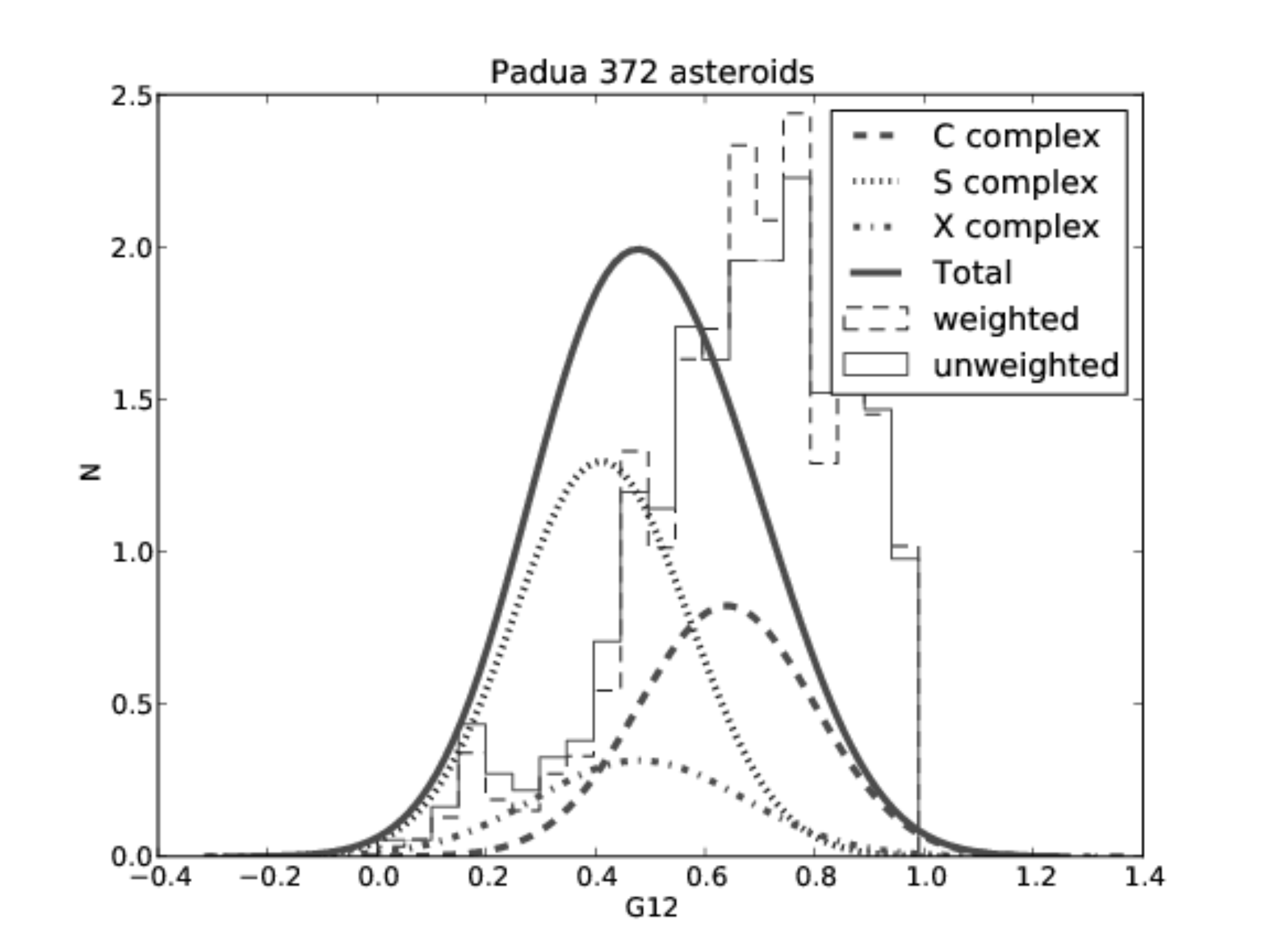} \\
\includegraphics[width=0.5 \textwidth]{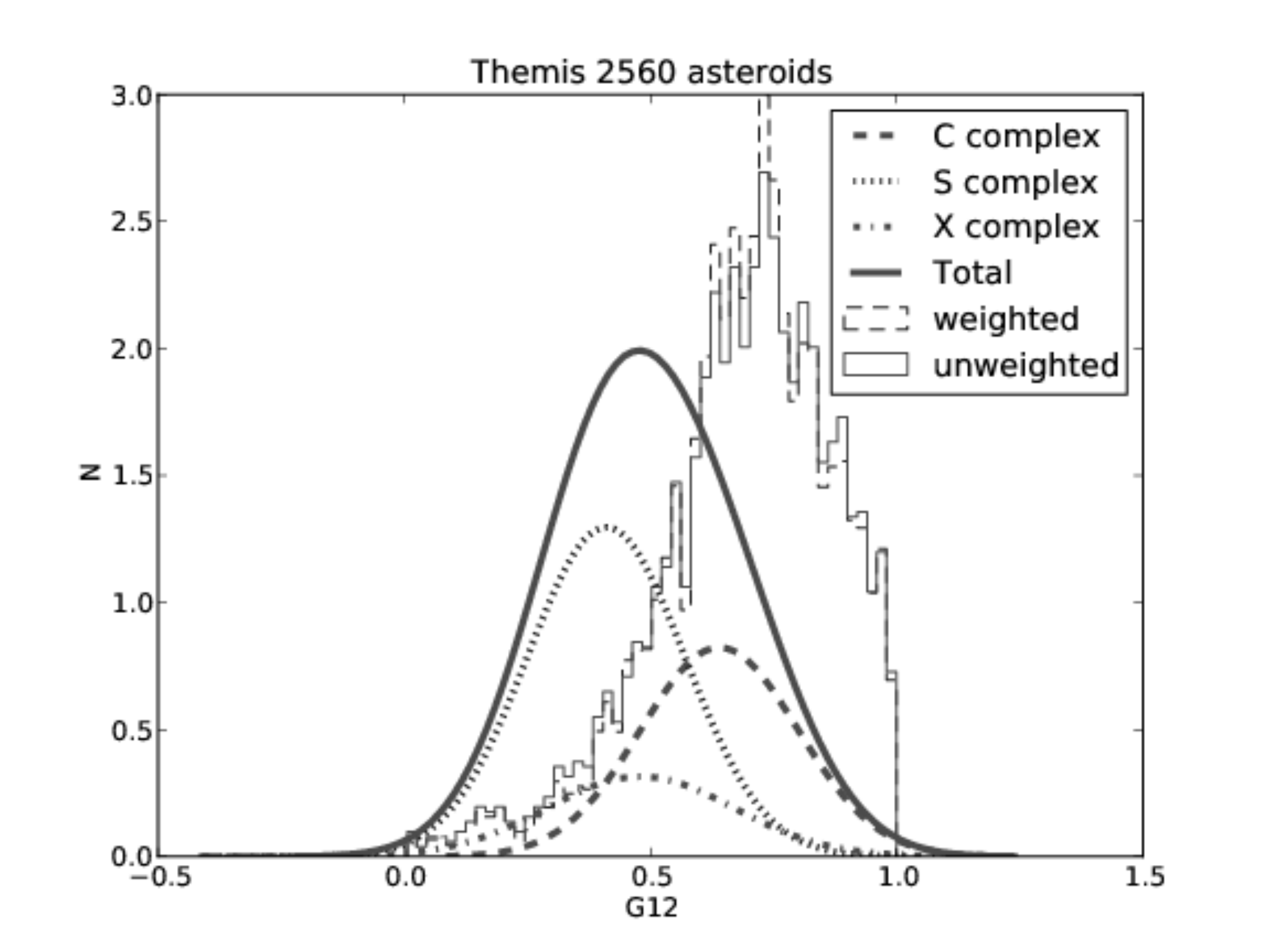} &
\includegraphics[width=0.5 \textwidth]{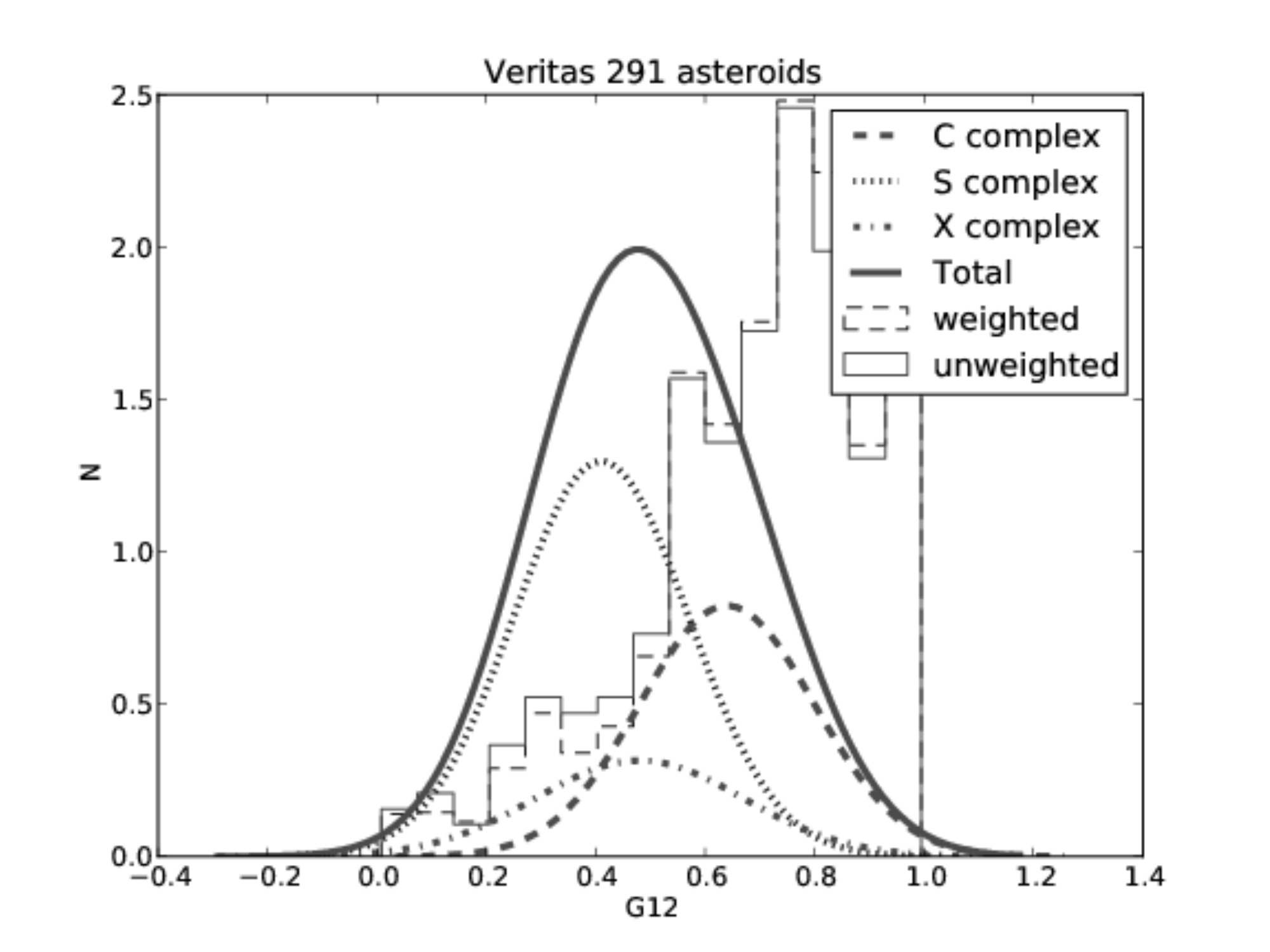} \\
\hline
\caption{The normalized $G_{12}$ distributions for asteroid families
  as listed in Table~\ref{mean}. The family classification is from PDS
  \citep{Nesvorny}. The dashed line indicates the distribution
  weighted with one over the sum of the absolute two-sided errors, and
  the solid line is the unweighted histogram. For comparison, we plot
  the a posteriori functions for the different taxonomic complexes 
  (the C complex is indicated by the thick dashed line, the S complex
  by the thick dotted line, and the X complex by the thick dash-dotted
  line) based on prior (2). }
\label{family}
\end{longtable}
\end{center}

Deciding on the taxonomic complex preponderance based on $G_{12}$ can
prove difficult and one should be careful in drawing conclusions when
the resulting probabilities for the different complexes are
similar. To pick asteroid families that show a preference in taxonomic
complex, we set the following requirements:
\begin{enumerate}
\item The minimum number of asteroids in the sample must be around 100
  or more.
\item The probability for preponderance must be the highest for all
  the assumed a priori probabilities. This is to make sure that the
  inference is driven by data and not by the a priori distribution.
\item The probability for the preponderant complex must be close to
  50\% or more for all the assumed a priori distributions.
\end{enumerate}

Table \ref{mean} lists the probabilities for taxonomic complex
preponderance, along with the family means and standard deviations of
the $G_{12}$ values. Several families have too few members to draw any
conclusions. Some of the families result in very similar taxonomic complex
preponderance probabilies for each complex and therefore no complex can
be indicated as dominating. For some families the computed probabilities
suggest different preponderant complex based on different a priori probabilies.
For those families no conclusions could be made. Several families, however 
show clear preference of taxonomic complex. Those include:

\begin{itemize}
\item {\bf (145) Adeona} (region II): The $G_{12}$ distribution for the
Adeona family contains 987 asteroids and is visibly shifted towards
high $G_{12}$ values. Based on the computed probabilities, the Adeona
family seems to be dominated by C complex objects: the C complex
probabilities for the family are 49\%, 55\%, and 55\% based on the a
priori distributions (1), (2), and (3), respectively, which agrees
with the literature. The computed C complex preponderance probabilities
are 20\% to 37\% higher than those for the other complexes.
Visual inspection of the histogram also suggests
that the majority of asteroids in this family must have come from the
C-complex distribution (see Fig.~\ref{family}).  The $G_{12}$
distribution for this family is smooth. In the literature, the Adeona
family has 12 members with known spectroscopic classification: 9 in
class Ch, 1 in C, 1 in X, 1 in D, and 1 in class Xk \citep{Mothe}.

\item {\bf (627) Charis} (region III): The Charis cluster seems to be
strongly C complex preponderant. C complex preponderance probabilities
are 45\%, 55\% and 55\% for the a priori distributions (1), (2), and
(3). The $G_{12}$ distribution is smooth and clearly shifted towards
large $G_{12}$ values. The profile of the $G_{12}$ distribution also
matches the $G_{12}$ distribution profile of the C complex.

\item {\bf (410) Chloris} (region II): The profile of the $G_{12}$
distribution for the Chloris cluster is similar to that of the Charis
family, also matching the profile of the $G_{12}$ distribution for the
C complex. The computed probabilities also indicate C complex
preponderance: they are 48\%, 59\%, and 55\% for the a priori
distributions (1), (2), and (3), and are about 20\% larger 
than for any other complex. This cluster has also been
spectroscopically characterized as C complex dominant by
\cite{BusThesis}.

\item {\bf (668) Dora} (region II): The Dora family is strongly C complex
dominated. The probabilities of C complex preponderance are 56\%,
67\%, and 63\% for the priori distributions (1), (2), and (3), and are
28--51\% higher than those for the S and X complexes. Also, the
$G_{12}$ distribution is smooth and matches better the C complex
distribution than the S or X complex distributions. In the literature,
Dora has 29 members with known spectra, all belonging to the C complex
(24 in class Ch, 4 in C, and 1 in class B) \citep{Mothe, BusThesis}.

\item {\bf (283) Emma} (region III): The $G_{12}$ distribution for the Emma
family is smooth and dominated by asteroids with large $G_{12}$
values. The probability of Emma being C complex preponderant is 52\%,
59\%, and 63\% for the a priori distributions (1), (2), and (3). These
probabilities are about 30\% larger than those for the S and X
complexes. Therefore, Emma can be classified as C complex
preponderant.

\item {\bf (1726) Hoffmeister} (region II): The Hoffmeister family is C
complex dominant. The C complex preponderance probability is 53\%,
63\%, and 59\% for the a priori distributions (1), (2), and (3), and
is about 25\% larger than those of the S or X complex
preponderance. There are 10 members of this family with known spectra:
8 in class C (4 in class C, 3 in Cb, and 1 in class B), 1 in Xc, and 1
in class Sa \citep{Mothe}. The $G_{12}$ distribution for the
Hoffmeister family is peculiar and steadily increasing towards larger
$G_{12}$ values.

\item {\bf (10) Hygiea} (region III): Similarly to the Hoffmeister family,
the Hygiea family is C complex preponderant. The C complex
preponderance probabilities are 51\%, 61\%, and 61\% for the a priori
distributions (1), (2), and (3). The probabilities for the other
complexes are about 20--40\% smaller.  Most of the asteroids in the
family are of class B (C complex) \citep{Mothe}.
  
\item {\bf (569) Misa} (region II): The Misa family is C complex
preponderant. The C complex preponderance probabilities are 54\%,
60\%, and 60\% for the a priori distributions (1), (2), and (3), and
are about 35\% higher than those of being S or X complex preponderant.

\item {\bf (845) Naema} (region III): The Naema family is C complex
preponderant. The C complex preponderance probabilities are 51\%,
58\%, and 62\% for the a priori distributions (1), (2), and (3), and
are about 20--40\% higher than those of being S or X complex
preponderant.

\item {\bf (128) Nemesis} (region II): The Nemesis family is C complex
preponderant. The C complex preponderance probabilities are 54\%,
64\%, and 60\% for the a priori distributions (1), (2), and (3), and
are about 25--35\% higher than those of being S or X complex
preponderant.

\item {\bf (363) Padua} (region II): The $G_{12}$ distribution for the Padua
family is shifted towards large values of $G_{12}$ and indicates C
complex preponderance.  The C complex preponderance probabilities are
52\%, 59\%, and 59\% for the a priori distributions (1), (2), and
(3). The Padua family has 9 members with spectral classification. Most
of them are X class asteroids (6 in class X and 1 in Xc), and there
are also 2 C class members \citep{BusThesis, Mothe}.

\item {\bf (24) Themis} (region III): The Themis family is C complex
preponderant which agrees with the literature analyses. The C complex
preponderance probabilities are 55\%, 62\%, and 66\% for the a priori
distributions (1), (2), and (3), and are about 30--50\% larger than
those of being S or X complex preponderant. In the literature, 43
Themis family asteroids have spectra available. The taxonomy of these
asteroids is homogeneous: there are 36 asteroids from the C complex (6
in class C, 17 in B, 5 in Ch, and 8 in class Cb) and 7 asteroids from the X
complex (5 in class X, 1 in Xc, and 1 in Xk) \citep{Mothe,
  Florczak1999}.

\item {\bf (490) Veritas} (region III): The Veritas family is C complex
preponderant which agrees with the literature analyses. The C complex
preponderance probabilities are 54\%, 61\%, and 65\% for the a priori
distributions (1), (2), and (3), and are about 25--50\% higher than
those of being S or X complex preponderant. In the literature, the
Veritas family has 8 members with known spectra, all of them belonging
to the C complex: 6 in class Ch, 1 in C, and 1 in class Cg
\citep{Mothe}.

\end{itemize}

For a number of families it was not possible to indicate preponderant taxonomic complex.
Out of those a particular case is the Nysa-Polana family, which shows
a clear differentiation into two separate regions.

\begin{figure}[htb]
   \centering
   \includegraphics[width= \textwidth]{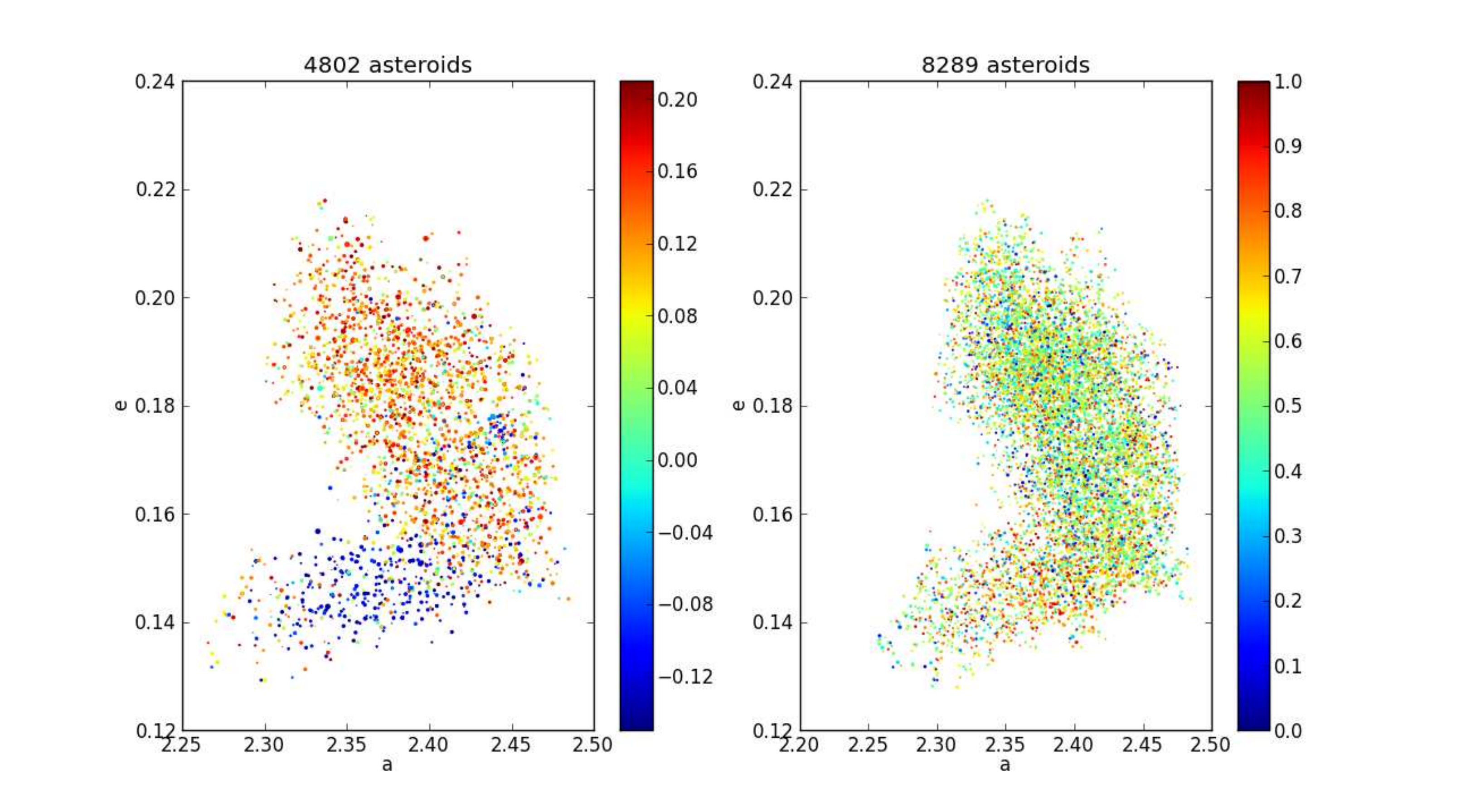} 
   \caption{Distribution of proper elements for asteroids in the
     Nysa-Polana region, color coded according to the SDSS $a^*$
     values (left) and the $G_{12}$ values (right).  The sizes of the
     points correspond to the errors in $G_{12}$ and SDSS $a^*$.}
   \label{NysaPolana}
\end{figure}

{\bf (44) Nysa - (142) Polana} (region I): The taxonomic preponderance
probabilities for the Nysa-Polana family are similar for all the
complexes, therefore no single complex can be indicated as
preponderant. Additionally, the different a priori distributions
result in differing preponderances. It has been previously suggested
\citep{CellinoNP} on the basis of spectral analysis that the
Nysa-Polana family is actually composed of two distinct families,
which is incompatible with the hypothesis of common origin. The first
one (Polana) was suggested to be composed of dark objects and the
second one (Mildred) of brighter S class asteroids. \cite{Parker2008}
has performed a statistical analysis and showed that, based on the
SDSS colors, it is possible to separate the Nysa-Polana region into
two families. in order to assess this suggestion, we plot the
distribution of proper elements (semimajor axis and eccentricity) of
the asteroids from the Nysa-Polana region in Fig.~\ref{NysaPolana},
color coded according to the $G_{12}$ values and the SDSS $a^*$ values
for comparison.

The two taxonomically different regions stand out both in $G_{12}$ and
in SDSS $a^*$. The sample size for the plot color-coded according to
the $G_{12}$ value is much larger than that color-coded with the SDSS
$a^*$ value. The $G_{12}$ plot suggests that there is much more
structure in the Nysa-Polana region and that there might be more than
the two main taxonomic groups present, or that they might be more
mixed. In their spectral analyses, \cite{CellinoNP} also found three
asteroids of class X, next to the 11 Tholen F class and 8 S class
asteroids. For reference, we list the $G_{12}$ values for the
main members of the Nysa-Polana region in Table~\ref{NP}. Generally,
it might turn out difficult to separate the two groups as they seem
quite strongly intermixed. We carried out a $k$-means clustering
operation for this region (with $k=2$). Clustering in the proper
elements and in the SDSS $a^*$ parameter gave, overall, the same
results as using the proper elements and the $G_{12}$ parameter.

\begin{table}[htb]
\centering
\caption{The $G_{12}$ parameters for the main members of Nysa-Polana family.}
\label{NP}
\vspace{1ex}
\begin{tabular}{llll}
Designation & H[mag] & $G_{12}$ & taxon \\ \hline
44 Nysa		& $+6.9^{+0.05}_{-0.05}$ 	& $+0.08^{+0.07}_{-0.07}$	& Xe \citep{pds} \\ 
142 Polana 	& $+10.20^{+0.013}_{-0.014}$ 	& $+0.69^{+0.09}_{-0.09}$	& B \citep{pds} \\
135 Hertha 	& $+8.13^{+0.01}_{+0.01}$ 	& $+0.32^{+0.08}_{-0.08}$	& Xk \citep{pds}\\
878 Mildred 	& $+14.51^{+0.03}_{-0.03}$ 	& $+0.79^{+0.17}_{-0.17}$	& \\ \hline
\end{tabular}
\end{table}                                           

Other families for which no conclusion could be made, but the shape of the $G_{12}$
distributions can still be discussed are:

\begin{itemize}
\item {\bf (847) Agnia} (region II, also called (125) Liberatrix): altogether 472
members are considered for the Agnia family, with a smooth but wide
$G_{12}$ distribution. The diverse $G_{12}$ values basically span
through the entire range of possible $G_{12}$ values. The decision
requirement (3) is not met for this family and therefore no definite
conclusions can be made. However, based on the large $G_{12}$ values
for many members of the family, we would suggest that the Agnia family
can contain large numbers of both S and C complex asteroids. In the
literature, the Agnia family has 15 members with known spectroscopic
taxonomy, all belonging to the S complex (8 in class Sq, 6 in S, and 1
in Sr) \citep{Mothe, BusThesis}.

\item {\bf (1128) Astrid} (region II): The $G_{12}$ values for 94 members of
the Astrid family result in a high probability of C complex
preponderance for the a priori distributions (1) and (3) and almost
equal probability of C and S complex preponderance for the a priori
distribution (2).  Accordingly, no conclusions can be made. In the
literature, the Astrid family has 5 spectrally characterized members,
all of which belong to the C complex (4 class C, 1 in Ch)
\citep{Mothe, BusThesis}.

\item {\bf (298) Baptistina} (region I): The $G_{12}$ distribution for the
Baptistina family is quite broad and results in similar probabilities
for all the complexes for the a priori distributions (1) and (2). For
the a priori distribution (3), the probability of S complex
preponderance is the largest.  In the literature, the Baptistina
family has 8 spectrally characterized members.  These asteroids tend
to have different spectral classifications: 1 in class Xc, 1 in X, 1
in C, 1 in L, 2 in S, 1 in V, and 1 in class A \citep{Mothe}. Due to
the lack of fulfillment of the requirements (2) and (3), we cannot
evaluate the taxonomic preponderance in this family.

\item {\bf (293) Brasilia} (region III): The $G_{12}$ distribution of the
Brasilia family is wide (spreading through the entire range of
possible $G_{12}$ values). The resulting preponderance probabilities
are similar for all the taxonomic complexes. In the literature, 4
members of the Brasilia family have known spectroscopic
classification: 2 in class X, 1 in C, and 1 in class Ch \citep{Mothe}.

\item {\bf (221) Eos} (region III): The Eos family has 92 members that have
taxonomic classification. There are 26 members in class T, 17 in D, 12
in K, 8 in Ld, 13 in Xk, 4 in Xc, 5 in X, 3 in L, 2 in S, 1 in C, and
1 class B. The family has an inhomogeneous taxonomy
\citep{Mothe}. This means that 43 asteroids originate from D complex,
25 from the S complex, 22 from the X complex, and 2 from the C
complex. In our treatment, we have decided to refrain from considering
complexes other than the three main ones, so indicating D complex
preponderance is not possible. The $G_{12}$ histogram for Eos is
shifted towards intermediate and large $G_{12}$ values, and is more
indicative of C complex rather than S or X complex preponderance.

\item {\bf (163) Erigone} (region I): The $G_{12}$ distribution for this
family is slightly shifted tpwards large $G_{12}$s. Erigone has 48\% and 54\%
probabilities of C complex preponderance based on the a priori
distributions (1) and (2), and a 48\% probability of S complex
preponderance based on the a priori distribution (3). Therefore, no
particular complex can be indicated as preponderant.

\item {\bf (15) Eunomia} (region II): The Eunomia family has a smooth
$G_{12}$ distribution with a profile matching the combined profile
of all complexes. The probabilities of Eunomia being S complex 
preponderant are 34\%, 42\%, and 42\% for the
a priori distributions (1), (2), and (3) and are the largest among the
different complexes, which agrees with the literature analyses. The
difference between the S and the other complex preponderance
probabilities are however only about 10\%. Generally, the
probabilities are below the required 50\%, so no complex can be
indicated as preponderant. In the literature, the Eunomia family has
43 members that have observed spectra, most members classified as
belonging to the S complex. There are 16 members in class S (including
(15) Eunomia), 2 in Sk, 10 in Sl, 1 in Sq, 7 in L, 4 in K, 1 in Cb, 1
in T, and 1 in class X \citep{Lazzaro1999, Mothe}.

\item {\bf (8) Flora} (region I): The $G_{12}$ distribution for the Flora
family is smooth with a mean at $G_{12}=0.53$. The probability of
Flora being S complex dominant is the largest and is 63\% for the a
priori distributions (2) and (3). Assuming a uniform a priori
distribution leads to almost equal taxonomic complex preponderance
probabilities. In the literature, Flora is considered S complex
preponderant \citep{Florczak}. However, due to the lack of fulfillment
of the decision requirements, we do not make a final conclusion on the
taxonomic preponderance in this family.

\item {\bf (1272) Gefion} (region II, also identified as (1) Ceres or (93)
Minerva): In the literature, the Gefion family has 35 members that
have spectral classification. Out of these asteroids, 31 belong to the
S complex (26 in class S, 2 in Sl, 2 in Sr, 1 in Sq, and 1 in L), 2
belong to the C complex (1 in class Cb class and 1 in Ch), and there
is 1 X class asteroid \citep{Mothe}. Our complex preponderance
probability computation results in similar probabilities for all three
complexes and is inconclusive for this family. The $G_{12}$
distribution for Gefion spreads through all the complexes and
is slightly shifted towards higher $G_{12}$ values.

\item {\bf (46) Hestia} (region II): The $G_{12}$ distribution for the
Hestia family is similar to that of the Gefion family. Thus, no
conclusions can be made as the probabilities of taxonomic
preponderance are comparable for all the complexes.

\item {\bf (3) Juno} (region II): In the case of the Juno family, the $G_{12}$
distribution is similar to the two previous families, 
the taxonomic complex preponderance probabilities are similar for all the
complexes. Therefore, no single complex can be indicated as
preponderant.

\item {\bf (832) Karin} (region III): For the Karin family, the
preponderance probabilities are similar for all the complexes for the
a priori distribution (1).  Therefore, no single complex can be
indicated as preponderant. For the a priori distributions (2) and (3),
the probabilities are discordant. The $G_12$ distribution for this
family is wide, whithout a clear prefference for any of the complexes.

\item {\bf (158) Koronis} (region III): In the literature, the Koronis
family has 31 members with spectral classification. There are 29
asteroids from the S complex (19 in class S, 1 in Sk, 3 in Sq, 2 in
Sa, and 4 in class K), 1 from class X, and 1 from class D. The spectra
of eight of these members have been analyzed by Binzel et al.  who
found a moderate spectral diversity among these objects
\citep{Binzel1993, Mothe}. In our computation, the Koronis family has
the highest chance of being C complex dominated. However, the
probability of being S dominant cannot be excluded as it is also quite
high. Due to the lack of fulfillment of the decision requirement (3),
we do not make a final conclusion on the taxonomic preponderance in
this family.

\item {\bf (3556) Lixiaohua} (region III): The taxonomic preponderance
probabilities for the Lixiaohua family are quite high for the C complex.
The probability of C complex preponderance are 50\%, 43\% and 59\% for
priors (1), (2), (3) respectively. However for prior (2) the probability
of this family being S complex dominated is 45\%. 
Therefore, no single complex can be indicated as preponderant.
$G_{12}$ distribution for this family spans though at the full range of
allowed $G_{12}$ and shows surplus of high $G_{12}$s.

\item {\bf (170) Maria} (region II): Even though the Maria family has the
largest probability of being S complex preponderant, which agrees with
the literature, no conclusions should be made as the probabilities of
the C and X complex preponderance are large . In the literature, the
Maria family has 16 members which have spectral classification, all
belonging to the S complex (4 in class S, 5 in L, 4 in Sl, 2 in K, and
1 class Sk) \citep{Mothe}. $G_{12}$ distrubution for this family matches
the combine $G_{12}$ profile from all the complexes.

\item {\bf (20) Massalia} (region I): The taxonomic preponderance
probabilities for the Massalia family are similar for all the
complexes. Therefore, no single complex can be indicated as
preponderant. Additionally, the three different a priori distributions
result in differing preponderant complexes. $G_{12}$ distribution
for Massalia family seems to be slightly shifted towards high $G_{12}$s.

\item {\bf (808) Merxia} (region II): The taxonomic preponderance
probabilities for the Merxia family are inconclusive as none of the
computed probabilities arises significantly above the rest. In the
literature, the Merxia family has 8 asteroids spectrally
characterized: 1 member belongs to the X complex while the remaining 7
members belong to the S complex (3 in class Sq, 2 in S, 1 in Sr, and 1
class Sl) \citep{Mothe, BusThesis}. $G_{12}$ distribution for this family
is wide and also matches the total $G_{12}$ distribution of all complexes combined.

\item {\bf (1644) Rafita} (region II): The taxonomic preponderance
probabilities for the Rafita family are similar for all the
complexes. Therefore, no single complex can be indicated as
preponderant. $G_{12}$ distribution for this family
is wide and matches the total $G_{12}$ distribution of all complexes combined.

\item {\bf (752) Sulamitis} (region I): The $G_{12}$ distribution for the
Sulamitis family is shifted towards large values of $G_{12}$,
indicating C complex predominance. The C complex preponderance
probabilities based on the a priori distributions (1) and (2) are
large (52 and 62\%). But based on the a priori distribution (3), the
preonderance probability is only 45\% and, therefore, we draw no final
conclusions.

\item {\bf (9506) Telramund} (region III): The taxonomic preponderance
probabilities for the Telramund family are similar for all the complexes.
Therefore, no single complex can be indicated as preponderant.
$G_{12}$ distribution for Teramund shows slight surplus of high $G_{12}$s.

\item {\bf (1400) Tirela} (region III): Tirela family also seems to be C
complex predominant with the $G_{12}$ distribution shifted
towardslarge $G_{12}$ values. However, due to the lack of fulfillment
of the decision criterion (3), we refrain from making a final
judgment.

\item {\bf (4) Vesta} (region I): The Vesta family is dominated by the V
complex asteroids. Here we are not considering V complex asteroids.
$G_{12}$ distribution for this family is wide and also matches the 
total $G_{12}$ distribution of all complexes combined.

\item {\bf (18466)} (region II): The taxonomic preponderance probabilities
for the family of asteroid (18466) are similar for all the
complexes. Therefore, no single complex can be indicated as
preponderant. $G_{12}$ distribution for this family is wide, matches the 
total $G_{12}$ distribution of all complexes combined and shows a slight
surplus of low to inermediate $G_{12}$s.

\end{itemize}

Number of families have to few members in our sample to be analyzed. Those include: 
(396) Aeolia (region II, 55 members), 
(656) Beagle (region III, 38 members), 
(606) Brang\"{a}ne (region II, 37 members), 
(302) Clarissa (region I, 41 members), 
(1270) Datura (region I, 4 members), 
(14627) Emilkowalski (region II, 2 members), 
(4652) Iannini (region II 30 members),  
(7353) Kazuya (region II, 12 members), 
(3815) K\"onig (region II, 58 members), 
(10811) Lau (region III, 6 members), 
(1892) Lucienne (region I, 37 members),
(137) Meliboea (region III, 40 members), 
(87) Sylvia (region III, 30 members), 
(1189) Terentia (region III, 7 members), 
(778) Theobalda (region III, 60 members), 
(18405) (region III, 24 members).
Even though conclusions for those could not be made, it is worth mentioning several of these families. The $G_{12}$ distribution for Datura and Iannini families are shifted towards small $G_{12}$ values and could be candidates for an S complex preponderant families. Asteroid (1270) Datura is spectraly identified as S class \citep{pds}. The $G_{12}$ distribution for the Theobalda family seems to be shifted towards C complex asteroids making it a candidate for C complex dominated. Asteroid (778) Theobalda is classified as F type \citep{pds}. C complex preponderance is also possible for Meliboea family, which also has been previously classified as C complex preponderant \citep{Mothe}. The C complex preponderance probabilities are 55\%, 62\% and 67\% for the a priori distributions (1), (2), and (3), and are about 25--40\% larger than those of being S or X complex preponderant. (137) Meliboea is spectrally classified as C class asteroid \citep{pds}.

Overall, the strict decision criteria requirements result in C complex
preponderance in the Adeona, Charis, Chloris, Dora, Emma, Hoffmeister,
Hygiea, Misa, Naema, Nemesis, Padua, Themis, and Veritas families. Out
of these, Adeona, Chloris, Dora, Hoffmester, Hygiea, Themis, and
Veritas have spectral classifications that indicate C complex
preponderance. Padua has only 9 members spectrally classified
(7 of X complex and 2 of C complex). The Charis, Emma, Misa,
Naema, and Nemesis families are yet to be spectrally classified.

There are no families that we can indicate as S or X complex
preponderant mainly because those two are more difficult to
separate. Also, there are no families which would have the
distribution clearly shifted towards small $G_{12}$ values and fullfil
our strict decision criteria.

\begin{figure}[htb]
\centering
\subfigure[Distribution of C, S, and X complex asteroids in the main belt.]{
   \includegraphics[width=0.45 \textwidth]{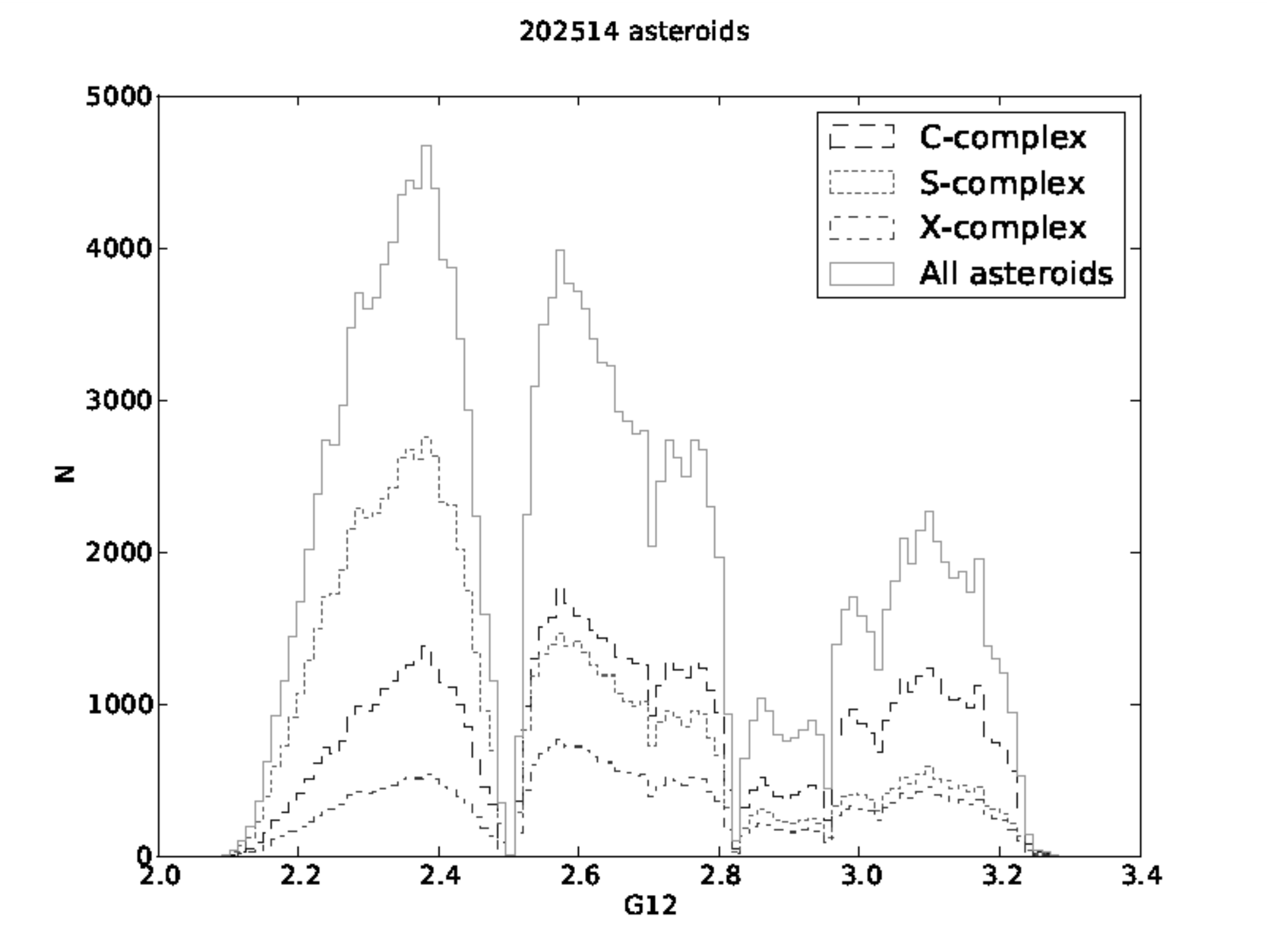} 
   \label{semiTaxa}
 }
 \subfigure[Weighted fraction of asteroid complexes thought the main belt.]{
   \includegraphics[width= 0.45 \textwidth]{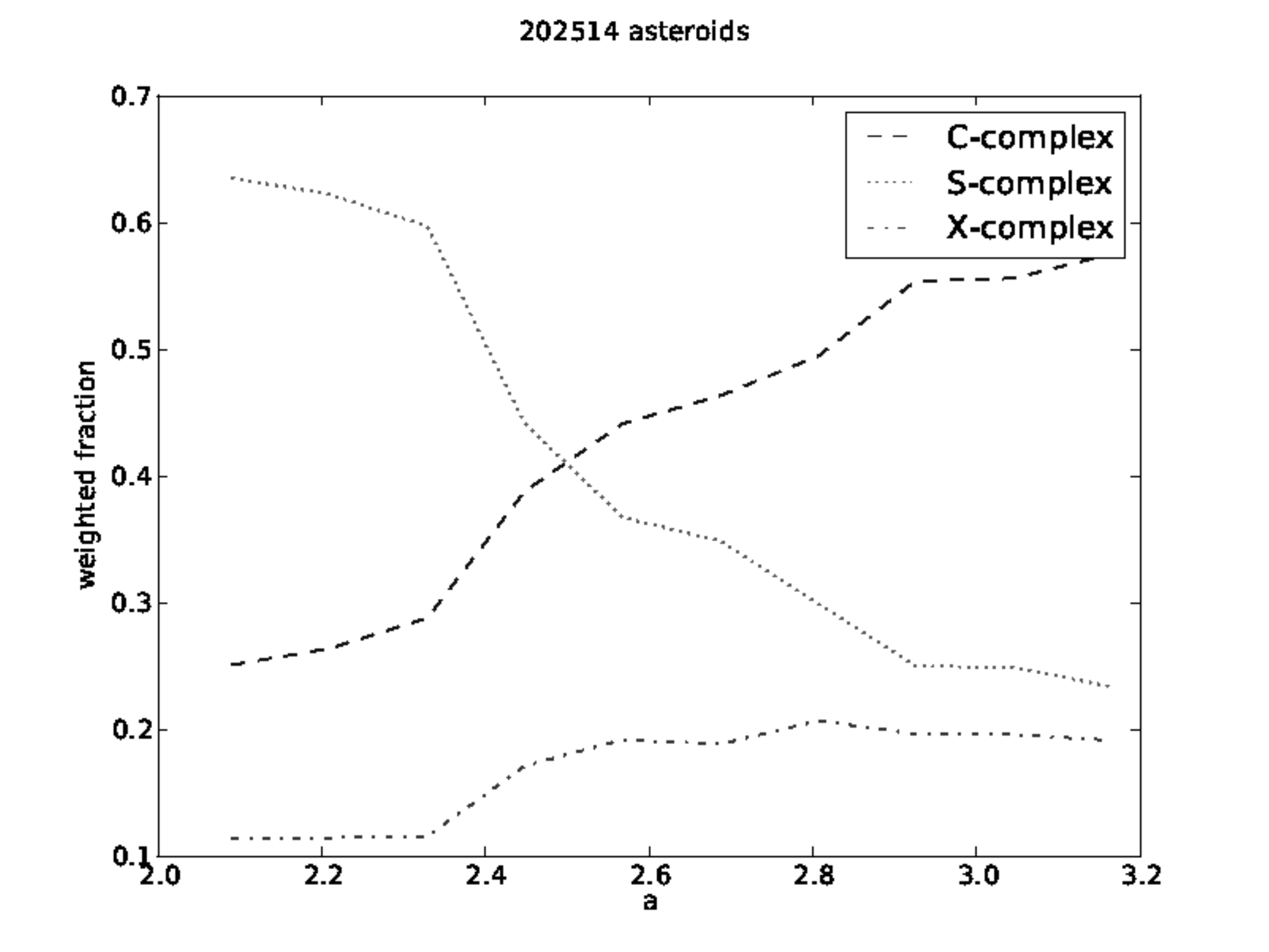} 
   \label{semiTaxaFraction}
 }
\label{semiTaxa_}
\caption{Composition of the main belt.}
\end{figure}

\begin{figure}[htb]
\centering
 \subfigure[With the artifact]{
   \includegraphics[width= 0.45 \textwidth]{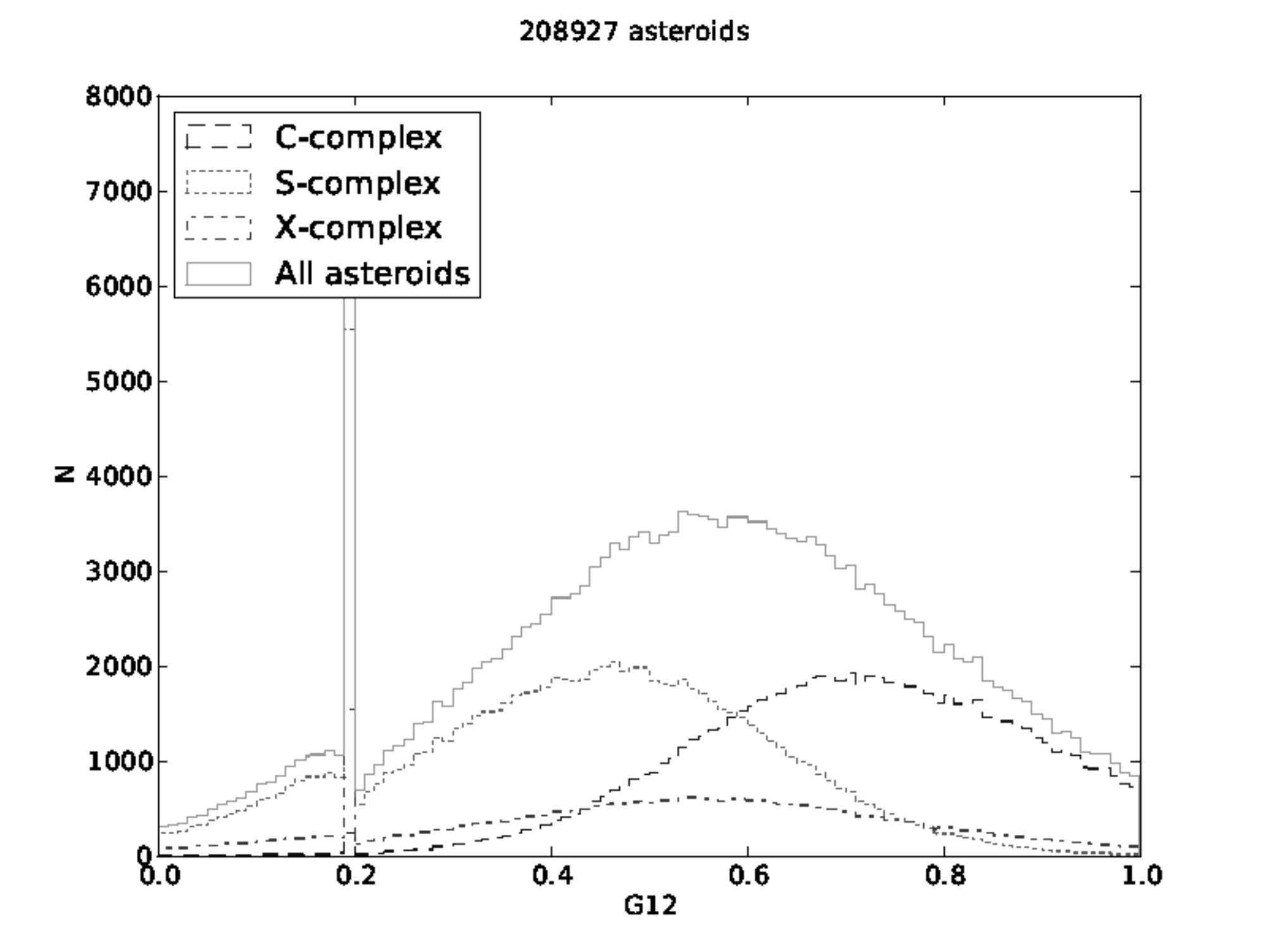} 
 }
 \subfigure[Artifact corrected]{
   \includegraphics[width=0.45 \textwidth]{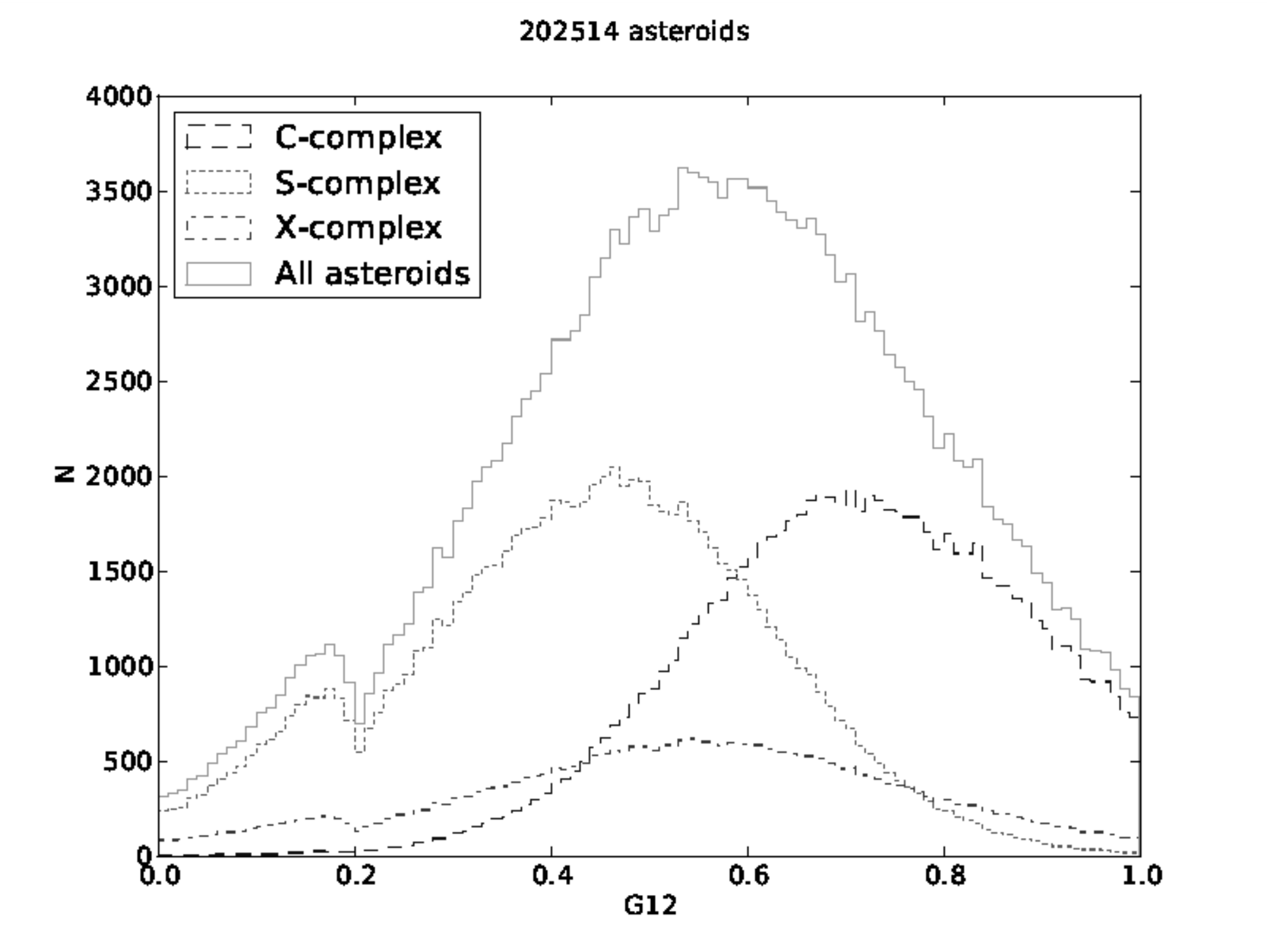} 
 }
\caption{Updated distribution of $G_{12}$ values in C, S and X complexes.}
\label{up}
\end{figure}

We have also computed taxonomic complex probability for all asteroids
having proper elements. Figure \ref{semiTaxa} shows the distribution
of C, S, and X complex asteroids in the main asteroid belt weighted
with the probabilities of belonging to the C, S, and X complexes.
Fig.~\ref{semiTaxaFraction} shows the weighted fraction of different
taxonomic complexes in the main belt. The overall distribution agrees
with the general view of more S complex asteroids in the inner main
belt and C complex asteroids dominating in the outer main belt
\citep[see, e.g.,][]{Gradie, Zellner, Thais, Yoshida, BusThesis}. On
the basis of the computed C, S, and X complex probabilities for each
asteroid, we modify the distributions of $G_{12}$ values for the C, S,
and X complexes in Fig.~\ref{up}. 
The gap at $G_{12}=0.2$ is related to the numerical function that
is 
used to derive the $H,\!G_{12}$ phase function, which is
nondifferentiable at $G_{12}=0.2$ \citep[for more details,
see][]{KM2010}. This causes many asteroids to end up at $G_{12}=0.2$
in least-squares fitting. To avoid the artificial peak at
$G_{12}=0.2$, we remove all asteroids with $G_{12}$ exactly equal to
$0.2$. However, from the dip at $G_{12}=0.2$, it is clear that some
valid solutions were thereby removed. Should the $H,\!G_{12}$ phase
function be revised, we recommend that the $G_1=G_1(G_{12})$,
$G_2=G_2(G_{12})$ functions are 
made differentiable. Figure~\ref{up} shows the updated $G_{12}$
distributions for the different complexes before and after correction
for the $G_{12}=0.2$ artifact.

\section{Conclusions} 
\label{concl}
We have analyzed the photometric parameter $G_{12}$ for all known
asteroids as well as $G_{12}$ distributions for asteroid families. We
have strengthened our previous finding of $G_{12}$ homogeneity in some
asteroid families and also confirmed a correlation between $G_{12}$
and taxonomy. $G_{12}$ could be potentially used in asteroid family
membership classification. We have further analyzed asteroid families
for C, S, or X complex preponderance. 

We conclude that, although
$G_{12}$ is related to surface properties, on its own it is mostly
insufficient to unambiguously assign taxonomic complex of individual
asteroids. Generally, the complex separation in the $G_{12}$
space is small. The $G_{12}$ distributions also overlap and,
generally, no definitive conclusions should be made for individual
objects based only on $G_{12}$ values. Rather probabilities of
belonging to a given complex can be computed. All classification
results for individual objects based on the $G_{12}$ values should 
be taken with caution and used rather to confirm previous results 
than to derive classifications based only on the current $G_{12}$ values.

In some cases $G_{12}$ values can, however, be an
indication of asteroid taxonomic complex. Particularly, the C complex
is the easiest to be separated, which was also confirmed by the high
success ratio in the testing. Accordingly $G_{12}$ distributions can 
be used in verifying taxonomic complex preponderance 
in some asteroid families. We found a preponderance of C complex asteroids in several
families. We compared our findings to the results available in the
literature, and concluded that, based on the $G_{12}$ distributions in
the families, we could confirm complex preponderance for several
families available. 

The $G_{12}$ values in conjunction with the SDSS colors could possibly
result in a better separation of the different taxonomic complexes. An
increased number of taxonomy classified asteroids and better quality
data potentially leading to better constrained $G_{12}$ values could
improve our knowledge of the $G_{12}$-taxonomy correlation. More
detailed knowledge of asteroids surface properties would also benefit
the classification problem. The Gaussian approximations of complex
distributions could be replaced by more sophisticated distributions,
and the a priori distributions for the Bayesian analysis could be
replaced by those deriving from debiased taxonomic
distributions. Particularly, a continuous debiased function describing
the fraction of different complexes throughout the belt could be used.

\section*{Acknowledgments} 
Research has been supported by the Magnus Ehrnrooth Foundation,
Academy of Finland (project No. $127461$), Lowell Observatory, and the
Spitzer Science Center.  We would like to thank Dr. Michael Thomas
Flanagan (University College London) for developing and maintaining
the Java Scientific Library, which we have used in the Asteroid Phase
Function Analyzer. DO thanks Berry Holl for help with Java plotters
and Saeid Zoonemat Kermani for valuable advice on Java applets. We
thank the Department of Physics of Northern Arizona University for CPU
time on its Javelina open cluster allocated for our computing.

\newpage
\bibliographystyle{elsarticle-harv}
\bibliography{biblio}

\end{document}